# Potential Biosignatures in Super-Earth Atmospheres II. Photochemical Responses

J. L. Grenfell[1], S. Gebauer[1], M. Godolt[1], K. Palczynski[1,*],
H. Rauer[1,2], J. Stock[2], P. v. Paris[2,#], R. Lehmann[3], and F. Selsis[4]

[1] Zentrum für Astronomie und Astrophysik, Technische Universität Berlin (TUB), Hardenbergstr. 36, 10623 Berlin, Germany, email: lee.grenfell@dlr.de
-to whom correspondance should be addressed

[2] Institut für Planetenforschung, Deutsches Zentrum für Luft- und Raumfahrt (DLR), Rutherford Str. 2, 12489 Berlin, Germany

[3] Alfred-Wegener-Institut für Polar- und Meeresforschung, Telegrafenberg A43, 14473 Potsdam, Germany

[4,#] Present address
(i) Univ. Bordeaux, LAB, UMR 5804, F-33270, Floirac, France
(ii) CNRS, LAB, UMR 5804, F-33270, Floirac, France

[*]Present address:
Helmholtz-Zentrum Berlin für Materialien und Energie GmbH, Hahn-Meitner-Platz 1, 14109 Berlin, Germany

Running title: Photochemistry Earth-like biosignatures

***Abstract:*** Spectral characterization of Super-Earth atmospheres for planets orbiting in the Habitable Zone of M-dwarf stars is a key focus in exoplanet science. A central challenge is to understand and predict the expected spectral signals of atmospheric biosignatures (species associated with life). Our work applies a global-mean radiative-convective-photochemical column model assuming a planet with an Earth-like biomass and planetary development. We investigated planets with gravities of 1g and 3g and a surface pressure of one bar around central stars with spectral classes from M0 to M7. The spectral signals of the calculated planetary scenarios have been presented by Rauer et al. (2011). The main motivation of the present work is to perform a deeper analysis of the chemical processes in the planetary atmospheres. We apply a diagnostic tool, the Pathway Analysis Program, to shed light on the photochemical pathways that form and destroy biosignature species. Ozone is a potential biosignature for complex- life. An important result of our analysis is a shift in the ozone photochemistry from mainly Chapman production (which dominates in Earth's stratosphere) to smog-dominated ozone production for planets in the Habitable Zone of cooler (M5-M7)-class dwarf stars. This result is associated with a lower energy flux in the UVB wavelength range from the central star, hence slower planetary atmospheric photolysis of molecular oxygen, which slows the Chapman ozone production. This is important for future atmospheric characterziation missions because it provides an indication of different chemical environments that can lead to very different responses of ozone, for example, cosmic rays. Nitrous oxide, a biosignature for simple bacterial life is favored for low stratospheric UV conditions, that is, on planets orbiting cooler stars. Transport of this species from its surface source to the stratosphere where it is destroyed can also be a key process. Comparing 1g with 3g scenarios, our analysis suggests it is important to include the effects of interactive chemistry.

## 1. Introduction

Understanding the photochemical responses of Super-Earth (SE) atmospheres in the Habitable Zone (HZ) of M-dwarf stars is a central goal of exoplanet science, since it is feasible that such environments may present the first opportunities to search for biosignature spectral signals. Gliese 581d (Mayor et al. 2009; Udry et al., 2007) is the first SE to be found that may orbit in the HZ of its M-dwarf star. Recently, initial constraints on the composition of hot transiting SEs such as CoRoT-7b (e.g., Guenther et al., 2011) and GJ1214b (e.g., Bean et al., 2011; Croll et al., 2011) have been discussed. Kepler 22b (Borucki et al., 2012) is the first transiting object found to occur in the HZ of a solar-type star; several Earth-sized objects have been found orbiting a cool M-dwarf (e.g., Muirhead et al., 2012) and detection of further SEs in the HZ is just beginning (e.g., Bonfils et al., 2013).

There exist a large number of possible parameters that could influence the abundances of possible biosignature species in hypothetical Earth-like atmospheres. Our motivation here is to take two parameters that are relatively well-known, namely, stellar class and planetary gravity, and perform a sensitivity study assuming an Earth-like biomass and development in order to determine their effect upon the photochemistry and climate, and hence the potential biosignatures. Other works (e.g., Segura et al., 2005; Grenfell et al., 2007) have also adopted this approach.

In this work, we analyzed the photochemical responses of key species from the same scenarios as the earlier work of Rauer et al. (2011) (hereafter Paper I), who analyzed spectral signals for Earth-like planets with gravities of 1g and 3g orbiting in the HZ of M-dwarf stars with classes from M0 to M7. In an earlier study, Segura et al. (2005) also discussed photochemical responses of (1g) Earth-like planets orbiting in the HZ of M-dwarf stars. They calculated enhanced abundances of methane ($CH_4$) (by about x100) and nitrous oxide ($N_2O$) (by about x5) compared with those of Earth related to the weaker UV emissions of M-dwarf stars. Their results also suggest a reduction in the ozone ($O_3$) column by up to about a factor of 7 compared with that of Earth, associated with weakened UV leading to a slowing in the $O_3$ photochemical source. This result was already broadly anticipated in the early 1990s (see Segura et al. 2005 and references therein). In the present study, we aimed to examine the nature of these photochemical responses in more depth. We applied a diagnostic tool termed the Pathway Analysis Program (PAP) written by Lehmann, (2004) to investigate the photochemical responses. PAP delivers unique information on chemical pathways of key species and has identified new chemical atmospheric pathways on Earth (Grenfell et al., 2006) and on Mars (Stock et al. 2012[a,b]). PAP is a key tool for understanding atmospheric sources and sinks of the biosignatures and related compounds. The usual mechanisms that operate in Earth's atmosphere (e.g., $O_3$ catalytic cycles etc.) are complex and may be very different for Earth-like planets orbiting M-dwarf scenarios, which is a good motivation for applying such a tool.

The primary driver of the photochemistry is the Top-of-Atmosphere (TOA) stellar flux, especially in the UVB and UVC regions, which weaken with decreasing effective stellar temperature. Therefore, we first analyzed the Ultra-Violet (UV) fluxes in our planetary atmospheres. Then, we focused on their influence on atmospheric ozone ($O_3$) since this is not only an important biosignature but also a key UVB absorber governing the abundances of other chemical species. We then investigated the biomarker $N_2O$, which is sensitive to UVB. Finally, we analyzed the photochemistry of $CH_4$ and water ($H_2O$) since these key greenhouse gases can influence surface habitability. We now present a brief overview of the photochemistry of the above four species.

### 1.1 Photochemistry of $O_3$

$O_3$ on Earth is a potential biosignature associated mainly with molecular oxygen ($O_2$), which arises mostly via photosynthesis. In Earth's atmosphere, about 90% (10%) of $O_3$ resides in the stratosphere (troposphere). *Production* of $O_3$ in the Earth's stratosphere occurs mainly via the Chapman mechanism (Chapman, 1930) via $O_2$ photolysis. Production of $O_3$ in the troposphere occurs mostly via the smog mechanism (Haagen-Smit, 1952), which requires volatile organic compounds (VOCs), nitrogen oxides, and Ultraviolet (UV).

*Destruction* of $O_3$ in the stratosphere proceeds mainly via catalytic cycles involving hydrogen-, nitrogen, or chlorine-oxides (e.g., Crutzen, 1970) (designated $HO_x$, $NO_x$, and $ClO_x$ respectively). These molecules can be stored in so-called reservoir species, the atmospheric distributions of which are reasonably well-defined for Earth (e.g., World Meteorological Organization (WMO) Report, 1995). Changes in, for example, temperature and/or UV can lead to the reservoirs releasing their $HO_x$-$NO_x$-$ClO_x$, associated with rapid stratospheric $O_3$ removal in sunlight. Destruction of $O_3$ in the troposphere occurs, for example, via wet and dry deposition and/or gas-phase removal via fast removal with, for example, $NO_x$.

$O_3$ can be formed abiotically in $CO_2$ atmospheres (e.g., Segura et al., 2007). $O_3$ layers (albeit very weak compared to that on Earth) have been documented on Mars (Fast et al., 2009) and on Venus (Montmessin et al., 2011), so caution is warranted when interpreting $O_3$ signals as indicative of biology or not (e.g., Selsis et al. 2002).

**1.2 Photochemistry of $N_2O$**

$N_2O$ is a biosignature produced almost exclusively on Earth from microbes in the soil as part of the nitrogen cycle (International Panel on Climate Change (IPCC, 2001)). Minor inorganic sources include, for example, the reaction of molecular nitrogen with electronically excited atomic oxygen: $N_2+O(^1D)+M \rightarrow N_2O+M$ (e.g. Estupiñan et al. 2002). Destruction of $N_2O$ occurs in the stratosphere mainly via photolysis or via removal with excited oxygen atoms.

**1.3 Photochemistry of $CH_4$ and Methyl choride ($CH_3Cl$)**

$CH_4$ is a strong greenhouse gas affecting climate and hence habitability. It is destroyed in the troposphere up to the mid-stratosphere mainly by oxidative degradation pathways with hydroxyl (OH) and in the upper stratosphere via photolysis. $CH_4$ is a possible indicator of life (bioindicator) but not a definite proof since this species (on Earth) has, in addition to biogenic sources, also some geological origins (IPCC, 2001).

$CH_3Cl$ on Earth has important biogenic sources associated with vegetation, although its source-sink budget and net anthropogenic ccontribution is not well known (Keppler et al. 2005). Like $CH_4$, its removal is controlled by reaction with OH, although the chlorine atom leads to increased reactivity (with an enhanced rate constant of about a factor 6 for this reaction) compared with $CH_4$.

## 1.4 Photochemistry of $H_2O$

Although not a biosignature, $H_2O$ is essential for life as we know it. Like $CH_4$, $H_2O$ is an efficient greenhouse gas. Production of $H_2O$ in Earth's stratosphere proceeds via $CH_4$ oxidation, whereas destruction of $H_2O$ occurs in the upper stratosphere via photolysis (World Meteorological Organisation (WMO), 1994). In the troposphere, $H_2O$ is subject to the hydrological cycle, including evaporation and condensation.

## 1.5 Key Questions

$O_3$ is formed on Earth in different ways, that is, via the smog mechanism (~10% on Earth) and the Chapman mechanism (~90%). How and why these values may change for different exoplanet scenarios is not well investigated, yet this is important information for predicting and interpreting spectra. A flaring M-dwarf star, for example, will induce a photochemical response creating $NO_x$, which destroys "Chapman"-produced $O_3$ but could actually enhance a "smog" $O_3$ signal.

$N_2O$ is destroyed via photolysis in the stratosphere by UVB radiation in the stratosphere, but its supply upwards from the surface is controlled by atmospheric transport and mixing. Models with fast upwards transport will ultimately lead to reduced $N_2O$ abundances since in the case of faster transport, the $N_2O$ molecules reach the altitudes of efficient destruction earlier, that is, the lifetime of $N_2O$ molecules is reduced, which (at a constant emission rate) leads to smaller $N_2O$ concentrations. To improve knowledge of potential $N_2O$ spectral signals in exoplanet environments, it is important to understand which processes (photochemistry or transport) dominate the abundance of $N_2O$ in different environments. For example, $N_2O$ on Earth is affected by both stratospheric UVB (which depends on, e.g., the solar spectra, radiative transfer, atmospheric photochemistry, etc.) as well as on tropospheric-to-stratospheric transport processes.

To begin to address such questions, we apply a new chemical diagnostic tool, the Pathway Analysis Program (PAP), which sheds unique light into the chemical pathways that control biosignature abundances.

## 2. Models and Scenarios

### 2.1 Models

The model details for the atmospheric coupled climate-chemistry column model and the theoretical spectral model have been described in Paper 1. Recent model updates include, for example, a new offline binning routine for calculating the input stellar spectra and a variable vertical atmospheric height in the model; more details were given by Rauer et al. (2011). The radiative-convective module is based on the work of Toon et al. (1989) for the shortwave region and RRTM (Rapid Radiative Transfer Module) for the thermal radiation. Since a main focus in this work is on photochemical effects, we will now provide a detailed description of the photochemical module. The model simulates 1D global-average, cloud-free conditions, although the effects of clouds were considered in a straightforward way by adjusting the surface albedo until the mean surface temperature of Earth (288 K) was attained for the Earth control run, as in earlier studies (Paper 1, Segura et al., 2003). The scheme solved the central chemical continuity equations by applying an implicit Euler solver that used the LU (Lower Upper) triangular matrix decomposition method with variable iterative stepping such that the stepsize was halved whenever the abundance of a long-lived species changed by more than 30% over a single step. The version used here employs chemical kinetic data from the Jet Propulsion Laboratory (JPL) Evaluation 14 (2003) report. The scheme includes the key inorganic gas-phase and photolytic chemical reactions commonly applied in Earth's atmosphere, that is, with hydrogen-, nitrogen, and chlorine-oxide reactions and their reservoirs. The scheme was considered to be converged when the relative change in concentration for any species in any layer changes by less than $10^{-4}$ over a chemical iteration that exceeded $10^5$s.

From a total of 55 chemical species, 34 were "long-lived," that is, the transport timescales are long compared with those of the photochemistry. Their concentrations were calculated by solving the full Jacobian matrix; 3 species, namely, $CO_2$, $N_2$, and $O_2$ were set to constant isoprofile values based on modern Earth, and the remainder of the species were "short-lived," that is, assumed to be in steady-state, and therefore calculated from the long-lived species. The steady-state assumption simplifies the numerical solution.

Surface biogenic and source gas fluxes for $CH_4$, (=531Tg/yr) $N_2O$ (=8.6 Tg N contained in $N_2O$ /yr) , CO (=1796Tg/yr) and $CH_3Cl$ (=3.4Tg/yr) were set such that for the Earth control run, Earth's modern-day concentrations were achieved at the surface – this procedure was commonly used in earlier approaches for Earth-like exoplanets (e.g., Paper 1, Segura et al., 2003). $H_2$ at the surface was removed with a constant deposition velocity of $7.7 \times 10^{-4}$ cm $s^{-1}$. Dry and wet deposition removal fluxes for other key species were included via molecular velocities and Henry's Law coefficients respectively. Volcanic fluxes of $SO_2$ and $H_2S$ were based on modern Earth. Tropospheric lightning sources of NO were based on the Earth lightning model of Chameides et al. (1977), assuming chemical equilibrium between $N_2$, $O_2$, and NO, a freeze-out temperature of 3500K and equilibrium constants taken from the Chemical Rubber Company (CRC) 1976 handbook. Modern Earth's atmosphere has ~44 lightning flashes $s^{-1}$ global mean (with flashes mainly generated over land in the tropics), which produces ~5Tg N in the form of NOx globally per year (Schumann and Huntreiser, 2007). Clearly, these values depend, for example, on atmospheric transport, convective activity, and the land-sea distribution, etc. for Earth-like exoplanets, which are not well-constrained parameters. At the model upper boundary, a constant, downwards (effusion) flux of CO and O is set, which represents the photolysis products of $CO_2$ that are formed above the model's upper lid.

Atmospheric mixing between the 64 vertical chemical layers was calculated via eddy diffusion constants (K in $cm^2$ $s^{-1}$), where log(K) varied from ~5.0 at the surface, decreased to a minimum value of ~3.6 at ~16km, and then increased to ~5.7 at the model upper boundary.

Photolysis rates included the major absorbers, including important (E)UV absorbers such as $O_2$, $CO_2$, $H_2O$, $O_3$, NO, $CH_4$, and $SO_2$. The $O_2$ photolysis absorbtion coefficients were calculated with the mean exponential sums method. The $O_3$ coefficients included the Hartley-Huggins T-dependence based on data measured at 203K and 273K (and linearly interpolated between). Species that photolyze in the UVB that are relevant for $O_3$ destruction were also included, for example, nitric acid ($HNO_3$) photolysis was included – this is important for NOx release. Finally, weakly bound species that photolyze in the UVA/visible region, for example, $NO_3$, $N_2O_5$ were included. Photolysis rates were calculated based on insolation fluxes from the delta two-stream module (Toon et al. 1989). One hundred eight wavelength intervals were included from (175.4-855) nm in the UV and visible, nine intervals in the EUV from (130-175) nm, and one Lyman-alpha interval at 121.6 nm. Rayleigh scattering for $N_2$, $O_2$, and $CO_2$ was included.

The Pathway Analysis Program (PAP) was developed by Lehmann (2004) and applied by Grenfell et al. (2006) to Earth's stratosphere and by Stock et al. (2012[a,b]) to the martian atmosphere. In the present work, it is applied to Super-Earth planetary atmospheres. The PAP algorithm identifies and quantifies chemical pathways in chemical systems. Starting with individual reactions as pathways, PAP constructs longer pathways step-by-step. To achieve this, short pathways already found are connected at so-called "branching point" species, whereby each pathway that forms a particular species is connected with each pathway that destroys it. Branching point species are chosen based on increasing lifetime with respect to the pathways constructed so far. In this work, all species with a chemical lifetime shorter than the chemical lifetime of the species being studied (i.e., the biosignatures $O_3$, $N_2O$, and the greenhouse gas $CH_4$) are treated as branching point species. Since in general the chemical lifetime of all species varies with altitude, the choice of branching point species adapts to the local chemical and physical conditions. A detailed description of the PAP algorithm is given by Lehmann (2004). To avoid a prohibitively long computational time, pathways with a rate smaller than a user-defined threshold (in the present study, $f_{min}=10^{-8}$ parts per billion by volume per second (ppbv/s)) are deleted. The chosen f_min = $10^{-8}$ ppbv/s is sufficient for finding the 5 dominant pathways (e.g., of $N_2O$, $CH_4$ loss) as shown in the main table (Appendix 1). Stock et al. (2012[a]) discussed the effect of varying this parameter. PAP calculates the chemical pathways by taking as input (i) a list of chemical species, (ii) chemical reactions, (iii) time-averaged concentrations and reaction rates, and (iv) concentration changes arising only from the gas-phase chemical reactions only (i.e., not including changes in abundance from, e.g., mixing, deposition, etc). PAP calculates as output the identified chemical pathways with their associated rates. Information from PAP is used to interpret chemical responses.

**2.2 Scenarios**

Here, we analyze the model scenarios described in Paper I. We considered planets with masses corresponding to 1g and 3g with Earth-like (i.e., $N_2$-$O_2$) atmospheres with Earth's source gas emissions and initial p,T, and abundance profiles as for modern Earth. There are currently no observational constraints for the surface pressure of SE planets. On the one hand, theoretical studies, for example, that of Elkins-Tanton and Seager (2008), have suggested a wide-range of possible atmospheric masses resulting from outgassing on SE planets, whereas on the other hand, for example, Stamenković et al. (2012), who included a pressure-depenence of viscosity in the mantle, suggested rather weak SE outgassing rates. Given the current uncertainties, we therefore assume 1 bar surface pressure to be comparable with Paper 1 and earlier studies and to compare with our 1g scenarios. Our modeled p, T, and chemical output profiles are calculated self-consistently for planets around different central M-dwarf stars in the HZ (with the Sun-Earth case for comparison). We explore an extensive parameter range, considering planets orbiting M-dwarf stellar classes from M0 to M7. This is neccessary because atmospheric chemistry-climate coupling is strongly non-linear and, hence, general results from one set of stellar classes (e.g., M0 to M4) cannot be simply extrapolated to other stellar classes (e.g., M5-M7) – instead each scenario has to be calculated separately. Mixing ratios for radiative species are fed back into the climate module, which calculates a new T, p profile, and this is again fed back into the chemistry module. This iterative process continues until T, p, and concentrations all converge. The planets are placed at an orbital distance from their star such that the total energy input at the TOA equals the modern Solar constant of 1366 $Wm^{-2}$ (see Paper 1 for the stellar input spectra used). In total, the following eleven scenarios were investigated:

| | |
|---|---|
| 1g Sun (run 1) | Earth |
| 3g Sun (run 2) | Super-Earth with three-times Earth's gravity (3g) orbiting the Sun |
| 1g M0 (run 3) | Earth-like planet (1g) orbiting M0 star |
| 3g M0 (run 4) | Super-Earth planet (3g) orbiting M0 star |
| 1g M4 (run 5) | Earth-like planet (1g) orbiting M4 star |
| 3g M4 (run 6) | Super-Earth planet (3g) orbiting M4 star |
| 1g ADL (run 7) | Earth-like planet (1g) orbiting active AD Leonis (ADL)* |
| 3g ADL (run 8) | Super-Earth (3g) orbiting active AD Leonis |
| 1g M5 (run 9) | Earth-like planet (1g) orbiting M5 star |
| 3g M5 (run 10) | Super-Earth (3g) orbiting M5 star |
| 1g M7 (run 11) | Earth-like planet (1g) orbiting M7 star |

* Segura et al. (2005) and Rauer et al. (2011) adopted a spectral class of 4.5 based on the SIMBAD database, whereas Hawley and Pettersen (1991) used a value of 3.5.

## 2.3 Planetary Radiation Environment

*Incoming Stellar fluxes* ($F_*$) – These are the primary driver of planetary atmospheric photochemistry, especially in the UVB and UVC range, and are also central to habitability for life as we know it on Earth. A significant proportion of cooler M-dwarfs like those considered in our work may be active emitters of UV from their chromospheres or/and transition regions (see e.g., Walkowicz et al., 2008, France et al., 2012 in press). This could have a considerable impact upon the planetary photochemistry, climate, and associated biosignatures. How efficiently the UV is absorbed throughout the atmospheric column is closely linked with the photochemical responses and, hence, determines the final abundances of the biosignature. We therefore start our analysis by investigating the planetary radiation environment. We discuss UV radiation at the TOA and at the planetary surface, and present a validation of surface UV based on Earth observations

*Planetary TOA Radiation Analysis* - We analyzed the planetary TOA $F_*$ in the UVA, UVB, and UVC wavelength range for the different stellar scenarios in the top model layer. UVA corresponds to the model wavelength intervals from (315-400) nm; UVB corresponds to (280-315) nm; UVC corresponds to (175.4-280) nm.

To be comparable with Paper 1, we approximated the TOA stellar spectra for the M0 to M7 M-dwarf stars as Planck functions (other than for the Sun, which is for solar mean conditions based on the work of Gueymard et al. (2004), and for AD-Leo, for which measured UV-spectra are available, see

Paper I). The approach used in Paper 1 and, therefore, in this study as well was to employ Planck curve spectra that correspond to quiet M-dwarf stars with little emitted UV fluxes. Recent results (Reiners et al., 2012) suggest that >90% of hotter (M0 to M2) M-dwarf stars sampled are *quiet*, whereas >50% of the cooler stars (M4 and cooler) are *active*. Clearly, we are well-aware that smooth Planck functions do not include, for example, enhanced Lyman-alpha and UVC features, etc. characteristic of cool M-dwarf stars that may have very active chromospheric and coronal regions. However, direct observations of stellar spectra for the cooler M-dwarf stars (M5-M7) in the critical wavelength range ($\lambda$<UVA) in our photolysis scheme are presently not available, and hence we prefer to adopt such a Planck –spectrum approach. Future work will study the effect of varying (E)UV characteristic emissions in the input spectra. Further, by comparing results from scenarios in which Planck curve spectra are used with those for active stars, we can isolate the photochemical effects in the planetary atmosphere of varying stellar activity. Firstly, to get an overview, Table 1 compares ratios of UV emission for our considered M-dwarf scenarios with the Sun.

Table 1: Ratios of UV radiation for our M-dwarf-star (M7) scenario compared with the Sun (upper row) and for ADL.

| Scenario | UVA | UVB | UVC |
|---|---|---|---|
| (M7/Sun) | $5.1 \times 10^{-3}$ | $1.2 \times 10^{-3}$ | $3.2 \times 10^{-4}$ |
| ($M_{active}$/Sun) | $1.2 \times 10^{-2}$ | $1.2 \times 10^{-2}$ | $6.5 \times 10^{-2}$ |

Table 1 (row 1) suggests that our cool (M7) M-dwarf would emit less than 1% of the UVA,

UVB, and UVC radiation compared with the Sun. Comparing row 2) the active AD Leo M-dwarf star with the Sun suggests that UVA, UVB, and UVC for the flaring star amount to only (1-7)% of the total Solar radiation

Figures (1a-1c) show the TOA UVA, UVB, and UVC net flux (W $m^{-2}$). Figure 1 shows an increase with increasing stellar effective temperature as expected. The active AD-Leo flaring case is an especially strong emitter of UV due to its extremely active chromosphere. Modeled TOA UVB flux for Earth (~18.3 $Wm^{-2}$) compare reasonably well with available observations (e.g. $16 \pm 3$ $Wm^{-2}$; Benestad, 2006).

*Planetary Surface Radiation* – In the chemistry module, the UVA and UVB net fluxes required for the photolysis scheme are calculated from the top layer downward via the twostream module with Rayleigh scattering. Figures (2a-2b) show UVA and UVB net flux ($Wm^{-2}$) at the planetary surface as calculated in the chemistry module of this work. UVC is essentially zero at the surface so is not shown in Figure 2, and similarly for Figure 3. Generally, Figures 2a and 2b show an increase in planetary surface UV radiation with higher stellar temperatures, as for the TOA cases shown in Figure 1.

*Comparison with Earth Surface UV Radiation* - Global satellite observations from 1992-1994 (Wang et al. 2000, their Figure 6b) suggest observed UVB surface radiation for Earth of ~1.4$Wm^{-2}$ for cloud-free conditions. By comparison, Figure 2b suggests that our model over-estimates this value, calculating 2.3 $Wm^{-2}$ UVB for the Earth control run. Uncertainties include, for example, our straightforward treatment of clouds whereby we adjust the surface albedo (see above) as well as the challenge of representing, for example, time-dependent and, for example, latitude-varying $O_3$ photochemistry and UV absorption in a global-averaged 1D model.

*Ratio of Surface to TOA UV Flux* – This ratio (R) is shown for the 1g and 3g cases in Figures 3a and 3b for UVA and UVB respectively. R is an inverse measure of the UV shielding of an atmosphere. Figure 3a suggests that UVA passes efficiently through the atmospheres considered, as expected, since most values of $R_{net,UVA}$ are >0.7. The UVA ratio is not greatly dependent on the stellar temperature.

Figure 3b shows as expected a much stronger atmospheric extinction of UVB than for the UVA wavelengths, and there is now a clear dependency on stellar temperature. Weaker overhead $O_3$ columns in the cool M-dwarf cases lead to a strong rise in the ratio in Figure 3b. For the 3g scenarios (circles), a lowering in the atmospheric column by a factor of three resulted in less UV shielding and a rise in the surface UV.

## 3. Chemical Analysis

Here, we first compare briefly previous results (Segura et al., 2005) reported in the literature. Then, we discuss the general trends in column abundances of the biosignatures and related key species. Finally, we discuss the chemical responses for the vertical profiles that were also shown in Paper I.

### 3.1 Column Biomarkers (1g planets)

**Column $O_3$** in Figure 4a (blue diamonds) mostly decreased with increasing star class (i.e., decreasing $T_{eff}$ of the star) related to less UVB, therefore there was a slowing in the photolysis of molecular $O_2$ and hence a slowing in the Chapman cycle, a major source of $O_3$. The $O_3$ profile responses are discussed in more detail in section 3.6. The column values are shown in appendix 1.

**Column $N_2O$** in Figure 4a (red squares) generally increased with increasing star class. The cooler stars emit less UVB, which suggests a slowing in the photolytic loss of $N_2O$ in the planetary atmosphere and hence an increase in its abundance.

**Column $CH_3Cl$** in Figure 4a (green triangles) generally increased with increasing star class due to less OH, its major sink (see OH analysis, Table 2). The response is comparable to $CH_4$ (discussed in next section), which has a similar photochemistry. Spectral features of $CH_3Cl$, however, were too weak to be evident in the calculations of Paper 1 despite the enhanced column amounts for the cooler stars.

### 3.2 Column Biosignatures (3g planets)

For the 3g planets, we assumed a constant surface pressure of 1 bar, which led to the total atmospheric column being reduced by a factor of three, as already mentioned (Figure 4b). The general trends for $O_3$ and $N_2O$ remain for the 3g scenarios, that is, mostly similar to the corresponding 1g scenarios already discussed, although the reduced total column resulted in a cooling of the lower atmosphere due to a weaker greenhouse effect, as we will show (see Paper I also).

The $N_2O$ 3g response is linked with enhanced UVB penetrating the reduced atmospheric column compared with 1g, which leads to more photolytic loss of $N_2O$. A transport effect also took place. For the 3g case (with its lower model lid due to less atmospheric mass and higher gravity), the upward tropospheric diffusion of $N_2O$ was faster, for example, by about 50% in the mid to upper troposphere than the 1g case. This meant that $N_2O$ for the 3g case could reach the stratosphere faster, where it would be rapidly photolyzed.

### 3.3 Column Greenhouse Gases (1g planets)

In this section, we discuss the planetary atmospheric column abundances of $CH_4$ and $H_2O$ since they have a major impact on temperature via the greenhouse effect. Vertical profiles will be discussed later and can also be found in Paper 1.

*$CH_4$ Column Response* – Since the only source of $CH_4$ in the model is fixed biomass surface emission, the $CH_4$ response for the various runs is controlled by the main atmospheric $CH_4$ sink, that is, removal via the hydroxyl (OH) radical. OH is affected by three main processes:

**OH Source(s)**: for example, $H_2O+O(^1D) \rightarrow 2OH$ (where $O(^1D)$ comes mainly from $O_3$ photolysis in the UV).

**OH Recyling** reactions in which $NO_x$ species can interconvert $HO_x$ (defined here as $OH+HO_2$) family members via, for example, $NO+HO_2 \rightarrow NO_2+OH$.

**OH Sinks,** for example, reaction with $CH_4$ and CO (see e.g. Grenfell et al., 1999 for an overview).

Figure 4c suggests a strong $CH_4$ (green diamonds) increase with decreasing effective stellar temperature. Cooler stars are weak UV emitters, which favors a slowing in the OH source reaction above. Note also that greenhouse warming by the enhanced $CH_4$ favors a damp troposphere (more evaporation) and, hence, all else being equal would favor actually *more* OH (via more $H_2O$, see source OH reaction above). This is an opposing process which our results suggest is not the dominant effect. So, for a given model, calculating accurately the *net* effect will depend, for example, on a good treatment of, for example, the hydrological cycle, which is challenging for a global column model. To aid in understanding the $CH_4$ response, which is controlled by OH, Table 2 summarizes the OH sources, sinks, recycling budget, and associated quantities.

Table 2: Modeled (lowest atmospheric layer) and observed (surface) global-mean key species abundances (molecules $cm^{-3}$) and reaction rates (molecules $cm^{-3}$ $s^{-1}$) affecting $CH_4$ (and $H_2O$) for various 1g scenarios. *From Lelieveld et al. (2002).

| Quantity | 1g Sun | 1g M0 | 1g M4 | 1g M5 | 1g M7 | 1g ADL |
|---|---|---|---|---|---|---|
| OH | $1.3x10^6$ **(Obs.=$1.1x10^6$)*** | $1.0x10^5$ | $4.0x10^3$ | $2.8x10^2$ | 7.0 | $1.3x10^2$ |
| **OH Source reaction** Rate$_{(O(1D)+H2O \rightarrow 2OH)}$ | $3.4x10^5$ | $1.3x10^5$ | $6.6x10^4$ | $2.2x10^4$ | $4.1x10^3$ | $1.1x10^4$ |
| **OH recycling reaction** Rate$_{(NO+HO2 \rightarrow NO2+OH)}$ | $2.3x10^5$ | $1.5x10^5$ | $7.1x10^4$ | $3.1x10^4$ | $1.0x10^4$ | $2.3x10^4$ |
| ($HO_2$/OH) | $2.1x10^2$ | $1.6x10^3$ | $2.7x10^4$ | $2.2x10^5$ | $3.1x10^6$ | $2.8x10^5$ |
| ($NO_2$/NO) | 2.6 | 16.2 | 56.1 | 98.4 | 132.1 | 84.8 |
| $O_3$ | $4.7x10^{11}$ | $6.0x10^{11}$ | $4.6x10^{11}$ | $3.2x10^{11}$ | $1.8x10^{11}$ | $3.0x10^{11}$ |

*OH Abundances* – Control run (1g Sun) OH abundances in Table 2 are within ~20% of global-mean observed OH proxies for Earth. Table 2 suggests a strong decrease in OH from left to right (i.e., for decreasing stellar effective temperature) especially for the M7 case.

*OH Source Reaction Rates* - The source reaction rate (Sun) in Table 2, that is, $O(^1D)+H_2O\rightarrow 2OH$, is about 12 times weaker than indicated by the Whalley et al. (2010) study, which investigates (Earth) clean-air, tropical northern-hemisphere daytime OH. The factor 12 difference reflects a lowering due to day-night averaging in our global mean model (which accounts for ~factor 2 of the difference in OH) and the fact that the Whalley study considered tropical conditions. Concentrations of the trace specie $O(^1D)$ in the control run (=6x10$^{-8}$ ppbv at 30km) compared reasonably well with Earth observations (~3x10$^{-8}$ ppbv, Brasseur and Solomon, 2005). Table 2 suggests that the source reaction rate decreases from left to right, which is consistent with the decrease in OH.

*OH Recycling Reaction Rates* - Our (Sun case) recycling reaction was comparable with that of the Whalley et al. (2010) study to within about 50%. Earlier (Earth) modeling studies, for example, that of Savage et al. (2001), suggest that the OH recycling reaction dominates the source reaction even in quite clean air-masses ($NO_x$ ~250pptv and below), which is somewhat in contrast to this and the Whalley study. In Table 2, the recycling reaction rates (like the source reaction) also decreases from left to right, which favors the decrease in OH, although the change in the source reaction is the stronger effect. For cooler stars, the recycling reaction becomes increasingly important compared with the source reaction, and it dominates for the ADL and M7 cases.

*$HO_x$ and $NO_x$ Ratios* – These ratios are sensitive markers of changes in $HO_x$ and $NO_x$ chemistry and hence affect, for example, $O_3$ cycles and $CH_4$. The ratios ($HO_2$/OH) and ($NO_2$/NO) in Table 2 increase strongly for the cooler stars. These ratios are strongly affected by the concentration of $O_3$, whose production via the Chapman mechanism (discussed in 3.5) weakens for the cooler stars. The ratios for the cooler stars are far from their "Earth" values, so the interactions between $HO_x$ and $NO_x$ are much perturbed. This is a hint that the usual mechanisms that operate on Earth (e.g., $O_3$ catalytic cycles etc.)

may be very different for the cooler star scenarios – a good motivation for applying PAP as already mentioned.

*Atmospheric response for AD Leo* – Although the 1g ADL scenario featured lower OH (Table 2) than for M5, ADL featured *lower* $CH_4$ (Paper I) than M5. The upper layers (>60km) of the 1g ADL run showed very rapid destruction of $CH_4$ via OH – about five times faster than for M5. This was consistent with the high Lyman-$\alpha$ output of ADL leading to faster $HO_x$ enhancement via $H_2O$ photolysis.

*Water Column Response* – Figure 4c suggests that the increased $CH_4$ columns (green diamonds), with decreasing stellar effective temperature generally (except for M7), lead to higher $H_2O$ columns (green squares). Generally, for the cooler star scenarios, (up to and including M5), more $CH_4$ greenhouse heating leads to more water evaporation in the troposphere, and in the stratosphere, faster $CH_4$ oxidation leads to faster $H_2O$ production. However, for the M7 case (Figure 4c), although $CH_4$ increased, surface temperature did not, which suggests a saturation in the $CH_4$ greenhouse from M5 to M7, where the lower atmosphere becomes optically thick at very high $CH_4$ abundances. Surface cooling from M5 to M7 is also seen in the temperature profiles in Paper 1 (their Figure 3).

### 3.4 Column Greenhouse Gases (3g planets)

**$CH_4$ and $H_2O$** - Figure 4d has a similar format to Figure 4c but instead shows results for the 3g (instead of 1g in 4c)scenarios. The basic response to decreasing the effective stellar temperature at 3g is similar to the 1g case, that is, results suggest a column rise in $CH_4$ and in $H_2O$ but with a drop-off in the latter for the cooler stars. To gain more insight into the effect of changing gravity, upon $CH_4$, Table 3 shows the ratio (1g/3g) of the $CH_4$ column and for the near-surface atmospheric OH abundance:

Table 3: Ratio (1g/3g) for the $CH_4$ atmospheric column and for near-surface OH (midpoint of lowermost gridbox) for the Sun compared with M-dwarf star scenarios.

| Quantity | Sun | M0 | M4 | M5 | ADL |
|---|---|---|---|---|---|
| $CH_4$_col_1g/$CH_4$_col_3g | 0.7 | 0.3 | 0.2 | 0.1 | 2.2 |
| OH_surf_1g/OH_surf_3g | 1.1 | 3.7 | 8.5 | 47.8 | 0.2 |

Without calculating interactive photochemistry, a passive tracer would undergo a column reduction by a factor of three from 1g to 3g, because at constant surface pressure, increasing gravity by a factor of three leads to column collapse and a reduction in the overhead column by the same factor as the increase in gravity. In Table 3, therefore, a hypothetical, passive tracer (with no chemistry) would have a value of exactly three. The actual (with chemistry) $CH_4$ column ratios (row 1), however, are all less than three. The reduction is consistent with faster chemical loss at 1g than at 3g. To investigate this further, OH ratios are shown in Table 3 (row 2). They mostly (except ADL) increase for the cooler stars, suggesting a lowering in the 3g OH abundances compared with the corresponding 1g cases for the cooler stars. This is consistent with faster chemical loss at 1g. The reduction in OH for the 3g scenarios implies that, for example, the increase in UVB due to weaker shielding of some 3g atmospheres (*favouring* OH production) is out-weighed by the (opposing) feedback where reduced greenhouse warming at 3g led to a drier troposphere (disfavoring OH which is produced via $O(^1D)+H_2O \rightarrow 2OH$).

This is confirmed by the water column (open circles in Figure 4d), which suggests that the 3g compared with 1g (Figure 4c) scenarios led to a weakening in the greenhouse effect and hence tropospheric cooling (as seen in Figure 2 of Paper 1) and a general lowering in the $H_2O$ column (due to more condensation) by around a factor of ten (Figure 4d) compared with the 1g case (Figure 4c). In general, however, note that responses in chemical abundances do not scale directly with the column reduction at 3g compared with 1g since the effects of, for example, photochemistry are important.

Figures (4e, 4f) show the ratios (1g column/3g column) for biosignature and greenhouse gases respectively. The main point is that the values can lie far from a value of three (which would be expected for a passive tracer). This shows that it is important to include the effects of interactive chemistry. For the biosignature $O_3$ there is some indication of an increase in the ratio shown in Figure 4e for the cooler stars, which will be the subject of future study. For $CH_3Cl$ (Figure 4e) and $CH_4$ (Figure 4f) (which both have similar OH removal chemistry), the trend is downward for the cooler stars. The $H_2O$ (Figure 4f) scenarios are relatively more damp (with values >3) than for a purely passive tracer. This suggests more efficient production of $H_2O$ from $CH_4$ for the cooler stars at 3g than at 1g, for example, due to more UV in the thinner, 3g atmospheres.

### 3.5 Column-Integrated Pathway Analysis Program (PAP) Results

Figure 5 shows output of $O_3$ cycles from the PAP. The cycles (divided into production and loss cycles) found have been quantified according to the rate of $O_3$ production or loss through each particular cycle expressed as a percentage of the total rate of production or loss found by PAP (see also description of Appendix 1 below). Values are integrated over the model vertical domain. PAP analyses were performed for each of the 64 vertical column model chemistry levels, and the column-integrated values are shown in Figure 5. The full cycles referred to in Figure 5 can be found in the Appendix.

**Sun PAP Analysis –** Figure 5 confirms the expected result for $O_3$ production, that is, the Chapman mechanism dominates over the smog mechanism. For $O_3$ destruction, the column model suggests strong $NO_x$ contributions in the lower stratosphere, although an Earth GCM study (Grenfell et al., 2006) suggests a strong $HO_x$ contribution there. This result could reflect the challenge of 1D models of capturing 3D variations in photochemistry. Also, the column model does not include industrial emissions unlike the Earth 3D model. The result should be explored in future comparisons between the column model and 3D runs.

**Column-Integrated $O_3$ (1g) Production -** Figure 5a suggests a change from a mainly Chapman-based $O_3$ production for the 1g Sun and the warmer 1g M-dwarf stars, switching to a slower, mainly smog-based $O_3$ production for the cooler stars (1g M5 and 1g M7). This was related to the decrease in UVB for the cooler star scenarios, since UVB is required to initiate the Chapman mechanism via photolysis of $O_2$.

**Column integrated $O_3$ (1g) Destruction** – Figure 5a also suggests that the classical $NO_x$ and $HO_x$ cycles (see also Figures 6 and 7) that operate mainly in the stratosphere were the most dominant $O_3$ loss pathways for the Sun and warmer M-dwarf scenarios. For the cooler stars scenarios, the enhanced CO concentrations led to a CO-oxidation cycle gaining in importance.

**Column $O_3$ (3g)** – Behavior at 3g (Figure 5b) was broadly similar to 1g, except at 3g *both* Chapman and smog were important $O_3$ producers for the M5 case (i.e., not just smog as in the 1g case). Weaker atmospheric UVB absorption led to more penetration of UVB and hence an increased role for Chapman in the layers below.

**Column-Integrated Results Table for $O_3$, $N_2O$ and $CH_4$** – Appendix (1a-1c) shows the integrated

column mean PAP output for $O_3$, $N_2O$, and $CH_4$ respectively. Shown are (i) the column integrated rates (CIR) (in molecules $cm^{-2}$ $s^{-1}$) for all pathways found by PAP ("Found_PAP"), (ii) the CIR for only the pathways shown in the Appendix ("Shown_PAP") (shown are either the 5 dominant pathways or the first pathways that together account for >90% of the total formation or loss of found_PAP, whichever condition is fulfilled first), and (iii) the CIR as calculated in the chemistry scheme of the atmospheric column model ("total_chem"). Percent values for a particular cycle show its individual rate as a percentage of Found_PAP.

Comparing these three CIR values, it can be seen that for the $O_3$ production, which is relatively straightforward, the pathways found by PAP can account very well for the rate calculated in the column model chemistry module. For the $O_3$ loss pathways, which are rather more complex than the production, PAP can still account for generally more than ~90% of the rate from the chemistry module. For the sometimes very complex $CH_4$ pathways, with the value of $f_{min}$ chosen for this study, PAP can account for only up to about 50% of the rate from the chemistry module. Further tests suggested that decreasing the PAP input parameter $f_{min}$ (the minimum considered flux, currently set to $10^{-8}$ ppbv $s^{-1}$ for all runs) leads to improvement, but the resulting complex $CH_4$ cycles are beyond the scope of this paper (see also 3.6.3). We now discuss the individual cycles for each scenario.

**$O_3$ Column-Integrated Pathways**

Chemical pathways for the 1g Sun scenario in Appendix 1a mostly compare well with established results for Earth as discussed above. Appendix 1a suggests that for the 1g M0 scenario - due to less stellar UVB emission compared with the Sun - the Chapman mechanism for producing $O_3$ is somewhat suppressed (89.2%) and a new CO sink ("CO oxidation 1", 7.4%) appears, since CO is abundant. For the 3g M0 scenario, results suggest that Chapman features more strongly (96.7%) in the thinner 3g atmosphere compared with the corresponding 1g case. $HO_x$ and $NO_x$ remain important chemical sinks for both the 3g and 1g cases. The active star (1g ADL) features a stronger Chapman contribution (97.2%) compared with 1g M0 since ADL is especially active in the UV, which is important for Chapman-initiation (via molecular oxygen) with only modest changes for the 3g ADL case. For cooler non-active stars (1g M5), large changes are apparent compared with the warmer star cases. Less UVB emission from the cool M5 star leads to a switch to smog-type $O_3$ production ("smog 1", 57.8%). As discussed, the atmosphere is abundant in $CH_4$ and CO. Thus, the "CO-oxidation-1" cycle is an important $O_3$ loss pathway (36.8%). For the (3g M5) case, the thinner total column at 3g compared with 1g leads to a rise in UV, which is consistent with more Chapman $O_3$ (47.8%) production than the 1g case (7.5%). For $O_3$ loss, a complex $CH_4$ oxidation pathway involving $CH_3OOH$ becomes important (46.8%), which is not evident at 1g. The changed UV environment leads to a modest rise in $HO_x$ in the upper troposphere at 3g. Finally, for the coolest M-dwarf case (1g M7), $O_3$ production occurs via numerous types of smog mechanisms involving the oxidation of different VOCs, for example, CO, HCHO, and $CH_3OOH$. CO smog cycles become a key means of producing $O_3$ especially for the cooler stars. Like $CH_4$, an important sink for CO is the reaction with OH. As discussed, weakening UV emissions for the cooler stars leads to less OH and therefore an enhanced abundance of CO. Near the surface, CO mixing ratios correspond to: 0.09 (Sun), 9.0 (M4), 64 (ADL), and 426 (M7) parts per million (ppm). $O_3$ loss also involves $NO_x$ cycles but also a smog mechanism ("smog 7") where $O_3$ is the net oxidant, which is consumed to oxidize $CH_4$ and a CO oxidation cycle.

Smog cycles have larger rates for the M5 and M7 scenarios than for the Sun and M0 scenarios. This is because the important smog 1 cycle (producing $O_3$) is in competition with the CO-oxidation 1 pathway (destroying $O_3$). At high $O_3$ concentrations (for the Sun and M0 scenarios), (i) the reaction $NO+O_3 \rightarrow NO_2+O_2$ shifts the $NO_x$ family to favor $NO_2$. The reduction in NO leads to a slowing in the key reaction $NO + HO_2 \rightarrow NO_2+OH$ and hence slows the smog 1 cycle. Also at high $O_3$ concentrations,

(ii) the reaction $HO_2$+$O_3$ → OH + $2O_2$ favors the CO-oxidation 1 pathway. These two effects together, favor large smog rates for the M5 and M7 scenarios. In summary, total vertically integrated $O_3$ production and loss rates for the 1g Sun (=$1.9\times10^{13}$ molecules $cm^{-2}$ $s^{-1}$) are 68 times larger than for the 1g M7 case (=$2.8\times10^{11}$ molecules $cm^{-2}$ $s^{-1}$), which illustrates the change in the dominance from the rather fast Chapman chemistry to the slower smog mechanism.

**$N_2O$ Column-Integrated Pathways**

The main result of the PAP is that loss pathways from the $N_2O$ "viewpoint" are non-catalytic for all scenarios. In other words, loss occurs mainly directly via photolysis, which can be calculated from the photolysis rate without performing a PAP analysis for $N_2O$. We therefore only show (Appendix 1b) one scenario as an illustration, that is, the Sun scenario, which confirms results measured for Earth, that is, ~95% loss via photolysis (i.e., the sum of the 4 cycles involving $N_2O$ photolysis in Appendix 1b), and ~5% loss via catalytic reaction with $O(^1D)$ is similar to observed values quoted for Earth (e.g., 90-95% photolytic loss, 5-10% via reaction with $O(^1D)$, IPCC Third Assessment Report, see discussion to Table 4.4). The PAP finds no formation pathways of $N_2O$ via inorganic reactions, as expected since these are insignificant compared with surface biogenic input. For the M-dwarf scenarios, photolysis similarly remained the main removal mechanism, and the overall column integrated rate of removal decreased by about a factor of two for the M7 compared with the Sun case since the cooler stars emit less UV.

**$CH_4$ Column-Integrated Pathways**

Appendix 1c shows the PAP output for $CH_4$. Results suggest a large number of complex removal pathways that oxidize $CH_4$. PAP found no in-situ production pathways, since there are no inorganic reactions in our model that produce $CH_4$ in the atmosphere. The net removal can involve either complete oxidation of $CH_4$ to its stable combustion products: $H_2O$ and $CO_2$ (as in the "oxidation $2O_2$-a" pathway for the 1g Sun scenario) but can also involve only partial oxidation, for example, to intermediate organic species such as formaldehyde (HCHO), for example, as in the "Oxidation $O_2$" pathway (1g M0). Clearly, more complete oxidation is favored in oxidizing environments, for example, damp atmospheres with strong UV where OH is abundant.

The choice of oxidant in the net reaction will depend on the central star's particular UVB radiation output and its ability to release, for example, $HO_x$, $O_x$, or $NO_x$ from their reservoirs in the planetary atmosphere. Importantly for $O_3$ photochemistry, there are $CH_4$ cycles in which $O_3$ itself is the oxidant in the net reaction (see e.g. net reaction for several cycles from the 3g Sun case). This is an example where $CH_4$-oxidation does not lead to the more familiar $O_3$ (smog) production, but to the reverse effect where $O_3$ is consumed. Many of the $CH_4$ pathways are $NO_x$-catalyzed, as on Earth, although this is not the case for all scenarios (e.g., pathway "$CH_3OOH$-d" (3g M5) does not include $NO_x$).

### 3.6 Altitude-Dependent PAP Results

In this section we will present PAP results from the same scenarios as the previous section. However, here we will discuss the contribution of the PAP cycles as profiles varying in the vertical.

#### 3.6.1 Vertical Changes in Ozone ($O_3$) Production and Loss Cycles

Figure 6 shows the altitude-dependent PAP results, comparing production and loss pathways for the Earth case (Figures 6a, 6b) with the M7 case (Figures 6c, 6d). Similarly, Figures 7a, 7b compare ADL (1g) with M5 (1g) (Figures 7c, d). In Figures 6 and 7, the logarithmic x-axis shows the rate of change of $O_3$ associated with a particular cycle found by PAP, in molecules $cm^{-3}$ $s^{-1}$. The black and white text labels on these Figures indicate the names of the $O_3$ pathways, which can be found in Appendix 1a. Note that the logarithmic x-axis where results are plotted cumulatively (meaning to estimate the contribution of a pathway at a particular height one must subtract its left-hand side x-axis boundary from its right-hand side x-axis boundary) in Figures 6 and 7 means that the pathways shown on the right-hand side of the Figure can make up a strong overall contribution to the net rate of change despite having only a thin section (relatively small area).

For the Earth results (Figures 6a, 6b), the $O_3$ production and loss rates output by PAP compare well with middle atmosphere $O_3$ budgets derived for Earth, see for example the work of Jucks et al. (1996), their Figure 4. The Earth results (Figure 6a) in the top model layer show an uppermost region of $O_3$ production (thin, blue stripe), which arose due to the single reaction: $O_2$+O($^3$P)+M→$O_3$+M. This is linked with the model's upper boundary condition, where a downward flux of CO and O($^3$P) is imposed. This is done to parameterize the effects of $CO_2$ photolysis (forming CO and O($^3$P)), which takes place above the model's lid, for example, above the mid mesosphere. The resulting enhanced O($^3$P) in the uppermost model layer favors the direct $O_3$ formation pathway found by PAP. The enhanced $O_3$ source was balanced by an increase in the photolysis rate of $O_3$, and therefore the abundance decreased smoothly with altitude as expected. The effect of varying the upper boundary will be the subject of future work. $NO_x$ loss cycles dominate (>60%) the Earth lower stratosphere; $HO_x$ cycles are more important in the upper stratosphere. For the 3g case (3g Sun), the $O_3$ production pathways are similar to those of Earth, but $HO_x$ destruction is stronger (~70%) in the lower stratosphere, which is consistent with more UV penetration (releasing $HO_x$ from its reservoirs) for the thinner (3g) atmospheric column compared with the 1g case. The enhanced tropospheric $HO_x$, which also stimulated the “CO oxidation 1” cycle, accounted for (30-50%) of tropospheric $O_3$ loss.

For the warm M-dwarf star scenarios (e.g., 1g M0) – here, like the control (1g Sun), smog 1 dominates 50-60% of the $O_3$ production in the troposphere (with 10-20% arising from $CH_4$ smog cycles). The influence of the smog mechanism extends to high altitudes (up to about 20km) compared with the Earth control (which extends up to about 16km). “Chapman 1” (Appendix 1a) dominated the stratosphere. $O_3$ loss was dominated by the “CO-oxidation 1” pathway (60-80%) in the troposphere, $NO_x$ loss pathways in the mid-stratosphere, and $HO_x$ loss pathways in the upper stratosphere. For the 3g case (3g M0), the “smog 1” pathway contributes ~70% of $O_3$ production in the troposphere with the ~(10-15%) remainder in the troposphere coming from $CH_4$ smog pathways. “Chapman 1” is dominant in the stratosphere, and “Chapman 2” is dominant in the uppermost layers (see discussion above for Earth run 1). $O_3$ loss, like the 1g case, was dominated by “CO-oxidation 1” pathway in the troposphere (~90%) with different $HO_x$ cycles important for loss in the upper levels.

In Figure 7, ADL $O_3$ photochemistry production (Figure 7a) is rather similar, for example, to the Earth control (1g Sun) case (Figure 6a) in that Chapman production dominates the stratosphere and smog in the troposphere. However, for the 1g M5 run, results are very different from what occurs on Earth, since $O_3$ production is now dominated by the smog mechanism through much of the atmosphere. For ADL, $O_3$ production occurred mostly via “Smog 1” (70-80%) in the troposphere, with various $CH_4$ smog pathways making up between 10-20% in this region. “Chapman 1” dominated the stratosphere. $O_3$ loss was again dominated by “CO-oxidation 1” in the troposphere (70-90%) with a variety of $HO_x$ cycles important for loss in the upper levels. Intense Lyman-$\alpha$ radiation favored some enhancement of $H_2O$ photolysis (hence more $O_3$ loss via $HO_x$) in the 1g ADL scenario compared to, for example, the Earth control (run 1), but the effect was quickly damped (in the uppermost ~2 model layers) and the overall change in $O_3$ was small. For the corresponding 3g case (3g ADL), $O_3$ production pathways did not change greatly with altitude compared with the 1g case. $O_3$ loss pathways were also rather similar to the 1g ADL case, with the "CO-oxidation 1" pathway for 3g ADL dominating the lower atmosphere.

The cooler stars (M5, M7) show significant changes in the $O_3$ photochemistry compared with the other M-dwarf scenarios. The rather weak UV radiation of these cooler stars means that Chapman chemistry (requiring UV to break the strong $O_2$ molecule) is now only significant (up to ~50% $O_3$ production) (1g M5) in the uppermost (>60km) altitudes. The "CO-smog 1" pathway, however, is now significant over all altitudes, accounting for 60% of $O_3$ production in the troposphere and about 30% in the upper atmosphere. A variety of $CH_4$ smog pathways make up most of the remaining $O_3$ production (1g M5). For $O_3$ loss, the "CO-oxidation 1" pathway is again significant (50-70%) in the lower half of the model domain, whereas a variety of $NO_x$ cycles are important in the upper regions. For the coolest star considered (1g M7), the $O_3$ abundance is determined by mainly CO and $CH_4$ oxidation. First, "classical" smog production - with OH as the oxidant (mainly CO smog 1 and various $CH_4$ oxidation pathways)--produce $O_3$ but, on the other hand, $O_3$ in the M7 scenario can also act as an oxidant in pathways that oxidise, for example, $CH_4$ and CO.

The M7 case (Figures 6c-6d) shows that the CO smog mechanism dominates the $O_3$ production, whereas the CO oxidation cycle and the classical $NO_x$ cycle dominate the $O_3$ loss. Near the surface, some direct removal of $O_3$ occurred via the reaction: $NO+O_3 \rightarrow O_2+NO_2$ (Figure 6d). On Earth, more $NO_x$ usually leads to more $O_3$ production via the smog mechanism; the direct removal reaction is, however, sometimes important at high $NO_x$ abundances, for example, in city centers. In our M7 scenario, which does not have industrial $NO_x$ emissions, an important source of lower atmosphere $NO_x$ is from lightning. For the cool M-dwarf 3g case (3g M5, not shown), the "CO-smog 1" and "Chapman 1" pathways make almost equal contributions to the $O_3$ production budget in the middle atmosphere. "Chapman 1" contributes up to ~80% of local production in the upper levels (where UV is abundant), whereas the smog mechanism contributes up to ~70% in the lower layers. The smog contribution has a minimum of ~20% local production near the cold trap, which is consistent with low temperatures and a rather low OH abundance. For the $O_3$ loss pathways, results suggest an increase in complex CO and $CH_4$ smog pathways that consume $O_3$.

**3.6.2 $N_2O$** – For all scenarios, non-catalytic photolytic removal (>90%) is the main loss mechanism in the stratosphere. Catalytic removal involving reaction with $O(^1D)$ makes up the remainder (occurring mostly in the mid to upper stratosphere) of the $N_2O$ loss.

**3.6.3 $CH_4$** – Results suggest that a large number of loss pathways occur near the cold trap. For example, at 16km (1g Sun), the $CH_4$ pathways found by PAP with the value of $f_{min}$ chosen in this study could account for only about 20% of the total $CH_4$ change calculated in the column model. Low OH abundances and cold temperatures in this region are consistent with rather slow oxidation and a resulting complex mix of only weakly oxidized organic species with individual pathway contributions lying below the PAP threshold criteria chosen for the present study, but whose net effect is important. For this study, the PAP detection threshold was set to $f_{min} = 10^{-8}$ ppbv/s. OH-initiated oxidation of $CH_4$ is more favored on the lower layers but with relatively more $O(^1D)$-initiated oxidation on the upper levels, where this species is more abundant. A test run (not shown) where the $f_{min}$ value is decreased to $10^{-9}$ ppbv/s was found to address the above problem, that is, PAP was then able to account, for example, three times more $CH_4$ net change (for the Earth run), though with a notable increase in the overall number of pathways, each with small contributions to net the overall chemical change, beyond the scope of our work.

**3.7 Comparison with previous studies** - Compared with the results of Segura et al. (2005), our results are similar for $N_2O$ and $CH_4$ within 10-20% for the inactive (e.g., M4) and active (ADL) cases. For $O_3$, our atmopsheric column amounts are ~40% thicker (=270DU) compared with the Segura et al. (2005)

(=164DU) value for the ADL case. This results from changes in our photochemical scheme, including, for example, the parameterization of the lower boundary flux of $H_2$, as discussed in Paper I. Also our stellar insolation corresponds to 1366Wm-2 at the TOA, whereas Segura et al. (2005) scaled their incoming spectrum to obtain a surface temperature of 288K.

## 4. Spectral Detectability of Biomarkers

**$O_3$** - Paper I shows that the detection of $O_3$ is challenging especially for M7. To better understand $O_3$ detectability, improved stellar spectra for the cooler stars in the (E)UV are desirable especially in the UVB and UVC, where $O_3$ responds sensitively. M7 stars are statistically older and burn more slowly compared with lower spectral class stars, which means more developed convection zones and possibly larger differences in UV between flaring and quiet states for M7 than considered in our work (see Reiners et al. 2012).

**$N_2O$** - Clearly, the most favorable (planet/star) contrast ratios are associated with cool stars such as M7. However, Paper I shows that some spectral absorption features can be weakened, partly due to the large $CH_4$ abundance, which warms the stratosphere. The $N_2O$ spectral features were weak for the scenarios analyzed.

## 5. Conclusions

- The potential complex-life biosignature $O_3$ has a very different photochemistry for planets orbiting in the HZ of cool M-dwarf stars compared to that of Earth since the key mechanism switches from Chapman production to slower, smog production. Expected responses of $O_3$ produced by the smog cycle (which could be *favored* by increases in $HO_x$ and $NO_x$, e.g., by cosmic rays) could be very different than Chapman-produced $O_3$ (where $HO_x$ and $NO_x$ catalytically *destroy* $O_3$). This is important to consider when predicting and interpreting $O_3$ spectral features.

- The simple microbial-life biosignature $N_2O$ increases for the cooler stars, mostly related to weaker photolytic loss of $N_2O$ via weaker UVB in the middle atmosphere, as found too by earlier studies. In some cases, however, variations in transport become important. The amount of $N_2O$ in the middle atmosphere depends on the UV and on the rate at which this species can be transported upwards from the troposphere into the stratosphere where it is photolyzed.

- The greenhouse gas $CH_4$ responses and its removal pathways become complex especially for the cooler stars. $CH_4$ abundances generally increase for the cooler stars, a result also found in earlier studies, due to a lowering in OH, its main sink, which is associated mainly with a weakening in the main OH source reaction that requires UVB.

- The potential vegetation biosignature $CH_3Cl$ is enhanced in abundance by more than three orders of magnitude compared with the Earth run especially for cool M-star scenarios associated with low OH since reaction with this species is the main sink (see also $CH_4$ above). Like earlier studies, our results suggest that its spectral features are nevertheless very weak.

- Comparison of the 1g and 3g scenarios suggests that it is important to include interactive photochemistry when calculating biosignatures and greenhouse gas abundances. Reducing the mass of the atmosphere by, for example, a factor of three does not always lead to a reduction in, for example, biosignatures and greenhouse gases by a factor of three, due to interactive climate-photochemical effects.

## Acknowledgements

This research has been partly supported by the Helmholtz Gemeinschaft (HGF) through the HGF research alliance "Planetary Evolution and Life." F. Selsis and P. von Paris acknowledge support from the European Research Council (Starting Grant 209622: E3ARTHs).

**Figure Captions**

Figure 1: Planetary Global Mean Top-of-Atmosphere Incoming Radiation (W $m^{-2}$) for UVA (Figure 1a), UVB (Figure 1b) UVC (Figure 1c) for Earth's gravity.

Figure 2: As for Figure 1 but at the planetary surface for UVA (Figure 2a), UVB (Figure 2b) and UVC (Figure 2c) for Earth's gravity (1g).

Figure 3: As for Figure 1 but showing the Ratio (Surface/TOA) (at 1g and 3g) for UVA (Figure 3a), UVB (Figure 3b) and UVC (Figure 3c) Radiation (1g).

Figure 4a: Atmospheric columns (Dobson Units, DU) (1g) of biosignatures, ozone ($O_3$), nitrous oxide ($N_2O$) and methyl chloride ($CH_3Cl$).
Figure 4b: As for Figure 4a but for (3g) scenarios (same surface pressure (=1 bar) as 1g).
Figure 4c: As for Figure 4a but for column $CH_4$ (Dobson Units, DU) and $H_2O$ (DU).
Figure 4d: As for Figure 4c but for (3g) scenarios assume same surface pressure (1 bar) as 1g cases.
Figure 4e: Column (1g/3g) ratio for the same biosignatures as shown in Figure 4a.
Figure 4f: Column (1g/3g) ratio for the same greenhouse gases as shown in Figure 4c.

Figure 5: Pathway Analysis results for Global Mean Ozone Sources (Figure 5a) and Sinks (Figure 5b) for the Sun and for the M-dwarf star scenarios (1g) calculated by the Pathway Analysis Program. The pathways are shown in the PAP Tables in the Appendix.

Figure 6: Pathway Analysis results showing cumulative contribution of altitude-dependent $O_3$ production and loss pathways for the 1g Sun (Figures 6a, 6b) and for the 1g M7 scenarios (Figures 6c, 6d) plotted in the vertical and shown in (molecules $cm^{-3}$ $s^{-1}$). Black and white labels on the Figure correspond to the names of the individual cycles as shown in Appendix 1. Logarithmic x-axis tick labels correspond to factors of x2 x5 and x8 respectively. Note that the model vertical grid is variable depending on, e.g., greenhouse gas heating, which leads to an expansion in the vertical for cooler effective stellar temperatures.

Figure 7: As for Figure 6 but for the 1g ADL (Figures 7a, 7b) and for the 1g M5 scenarios (Figures 7c, 7d).

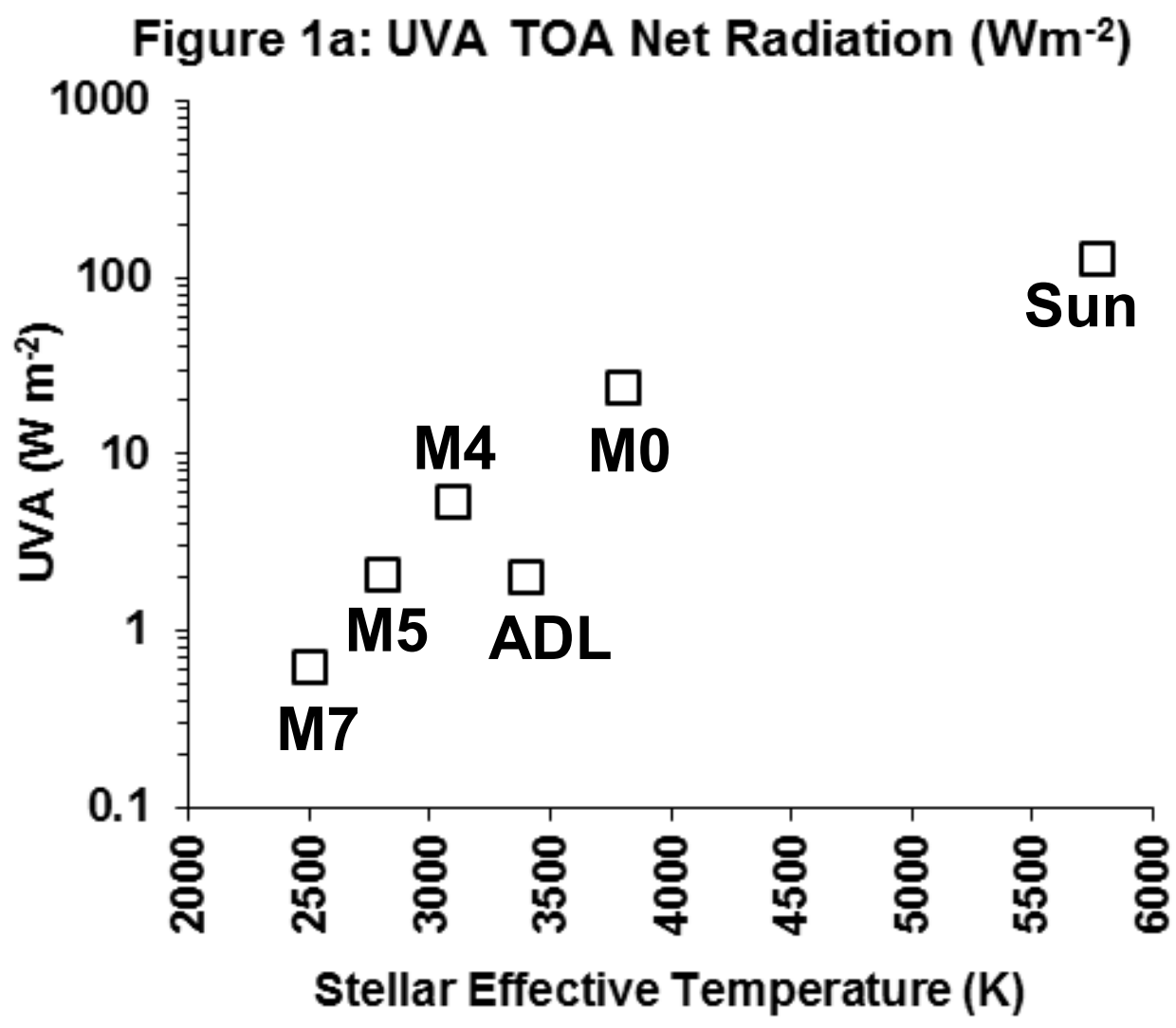

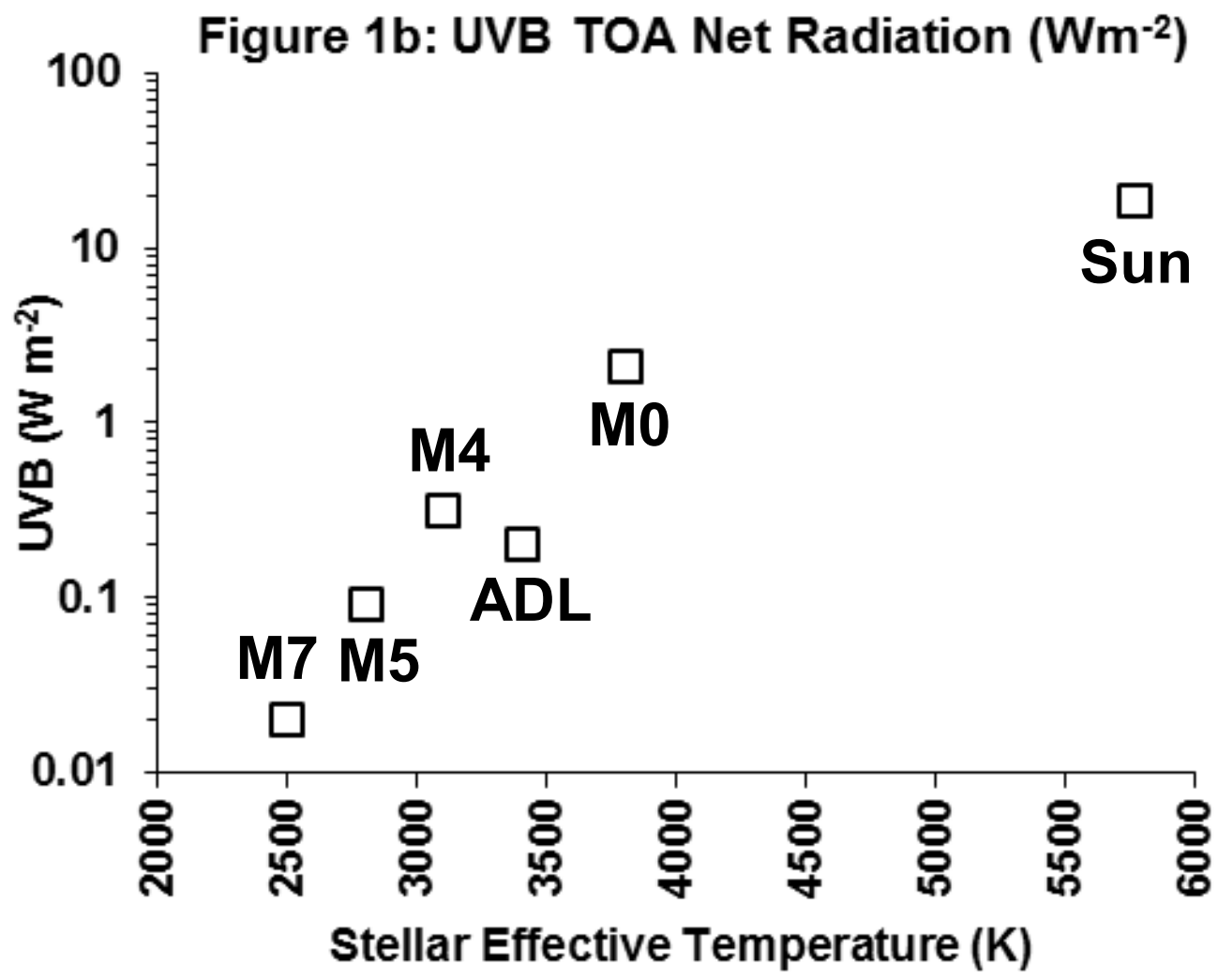

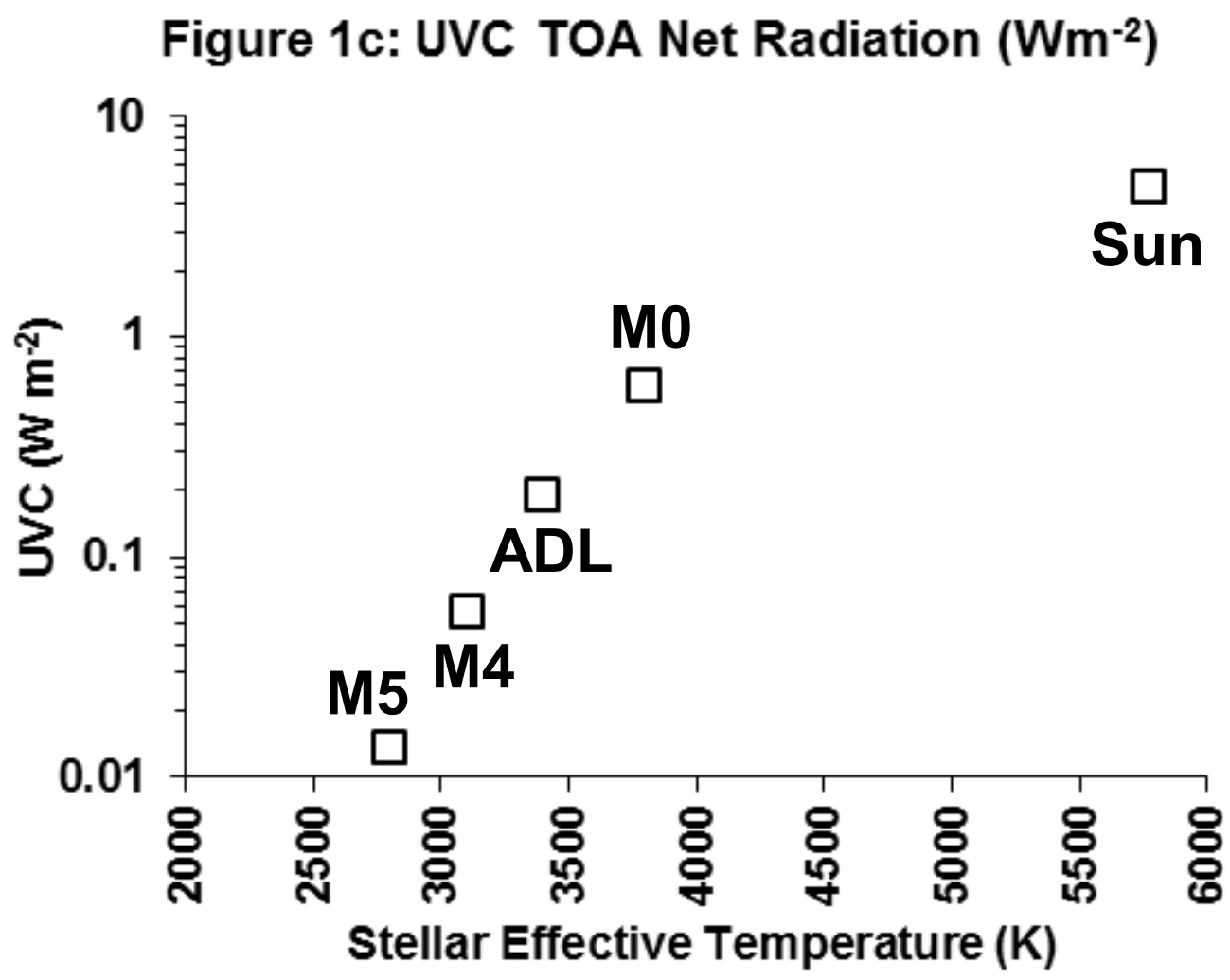

Figure 2a: 1g UVA Surface Net Radiation ($Wm^{-2}$)

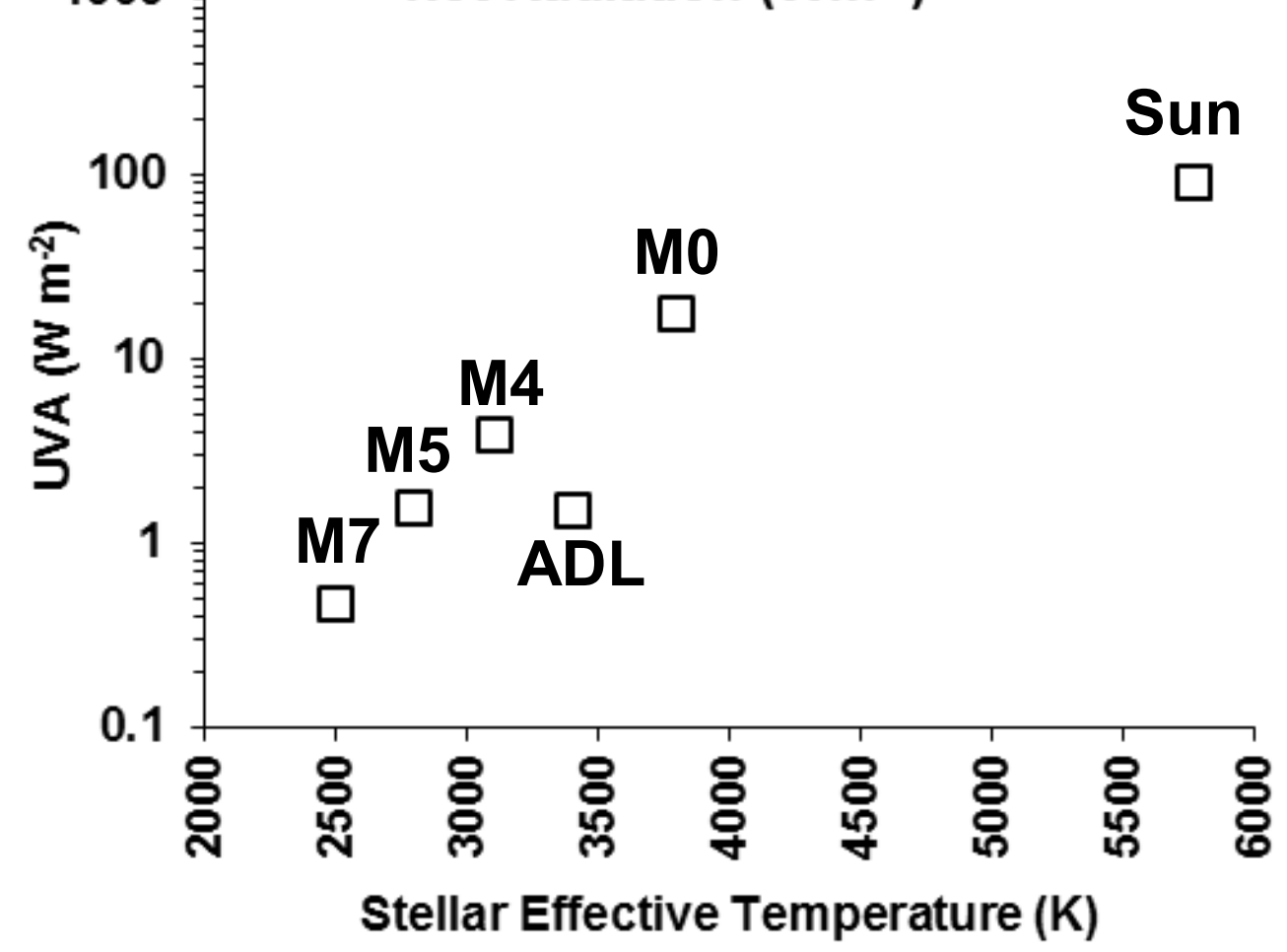

Figure 2b: 1g UVB Surface Net Radiation ($Wm^{-2}$)

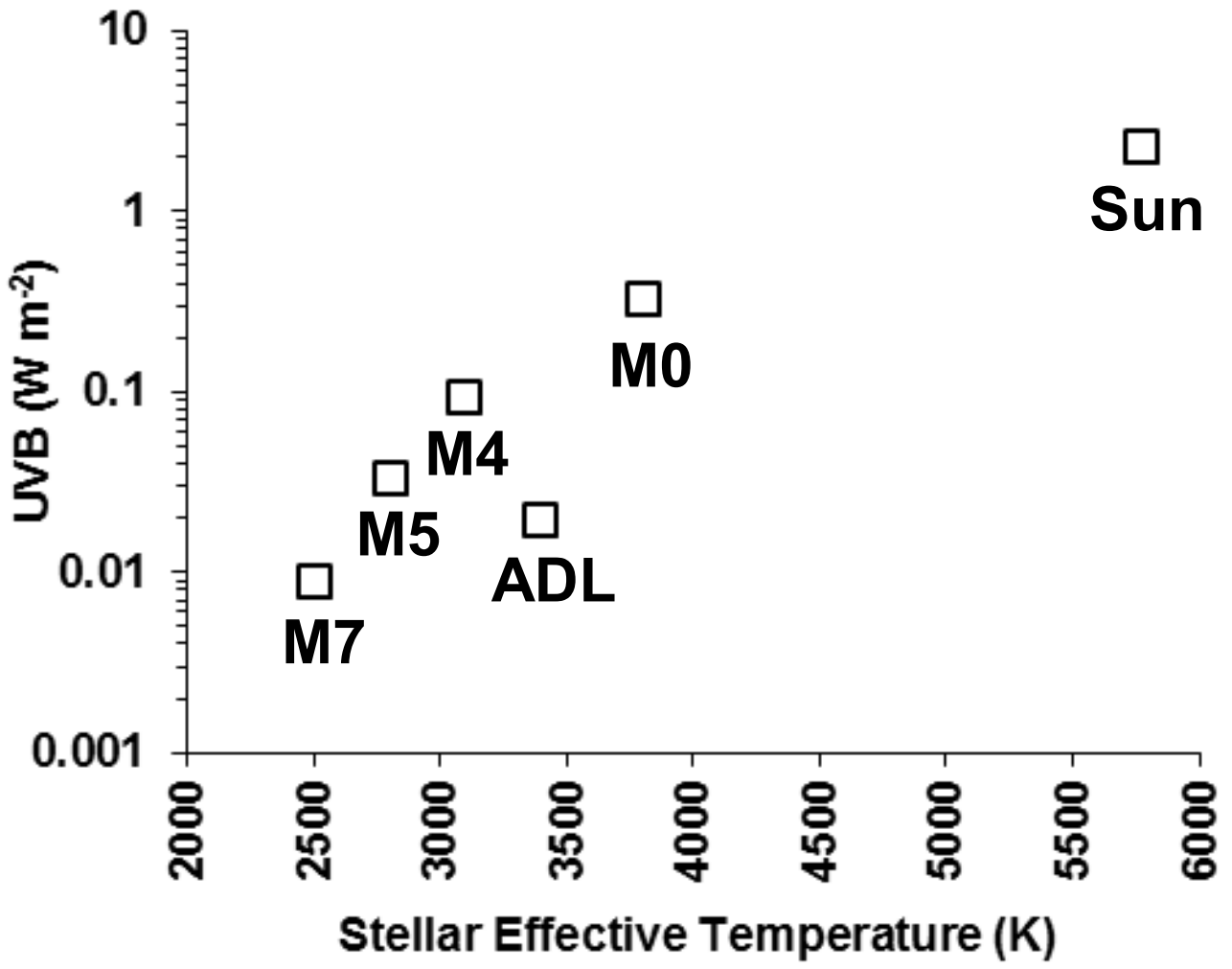

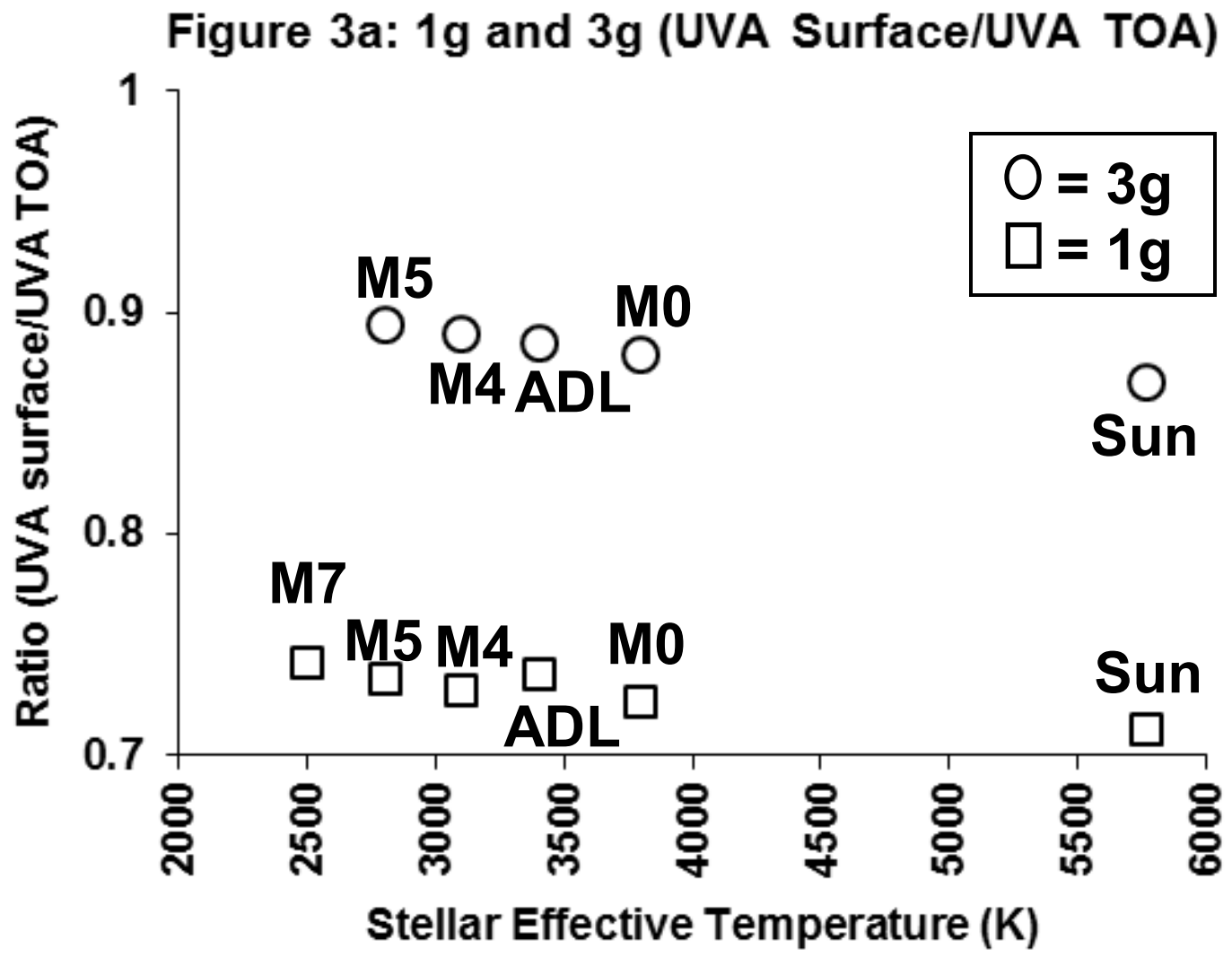
Figure 3a: 1g and 3g (UVA Surface/UVA TOA)
Ratio (UVA surface/UVA TOA)
Stellar Effective Temperature (K)
○ = 3g
□ = 1g
M5
M4
ADL
M0
Sun
M7
M5
M4
M0
ADL
Sun
1
0.9
0.8
0.7
2000
2500
3000
3500
4000
4500
5000
5500
6000

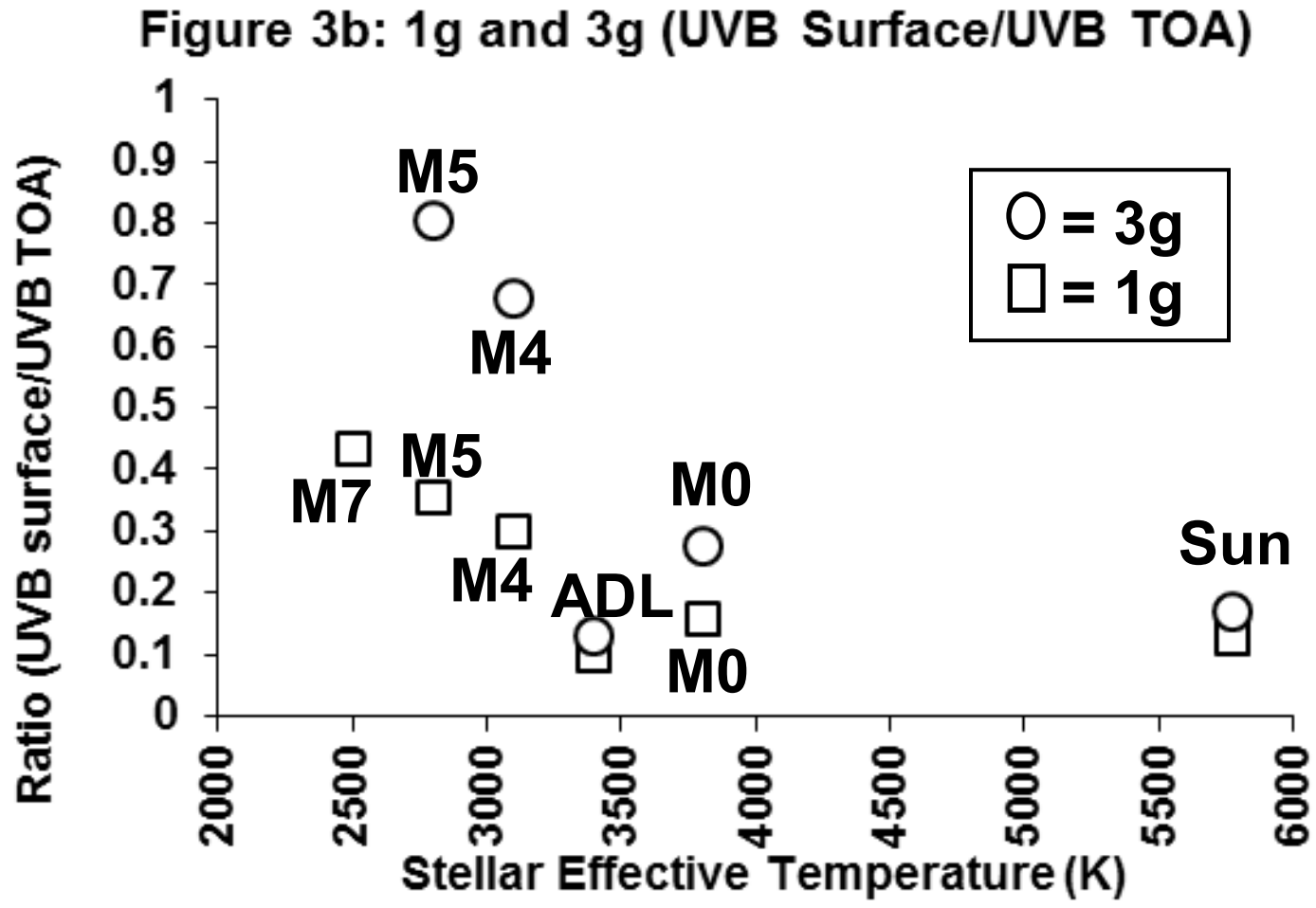
Figure 3b: 1g and 3g (UVB Surface/UVB TOA)
Ratio (UVB surface/UVB TOA)
Stellar Effective Temperature (K)
○ = 3g
□ = 1g
M5
M4
M5
M7
M4
ADL
M0
M0
Sun
1
0.9
0.8
0.7
0.6
0.5
0.4
0.3
0.2
0.1
0
2000
2500
3000
3500
4000
4500
5000
5500
6000

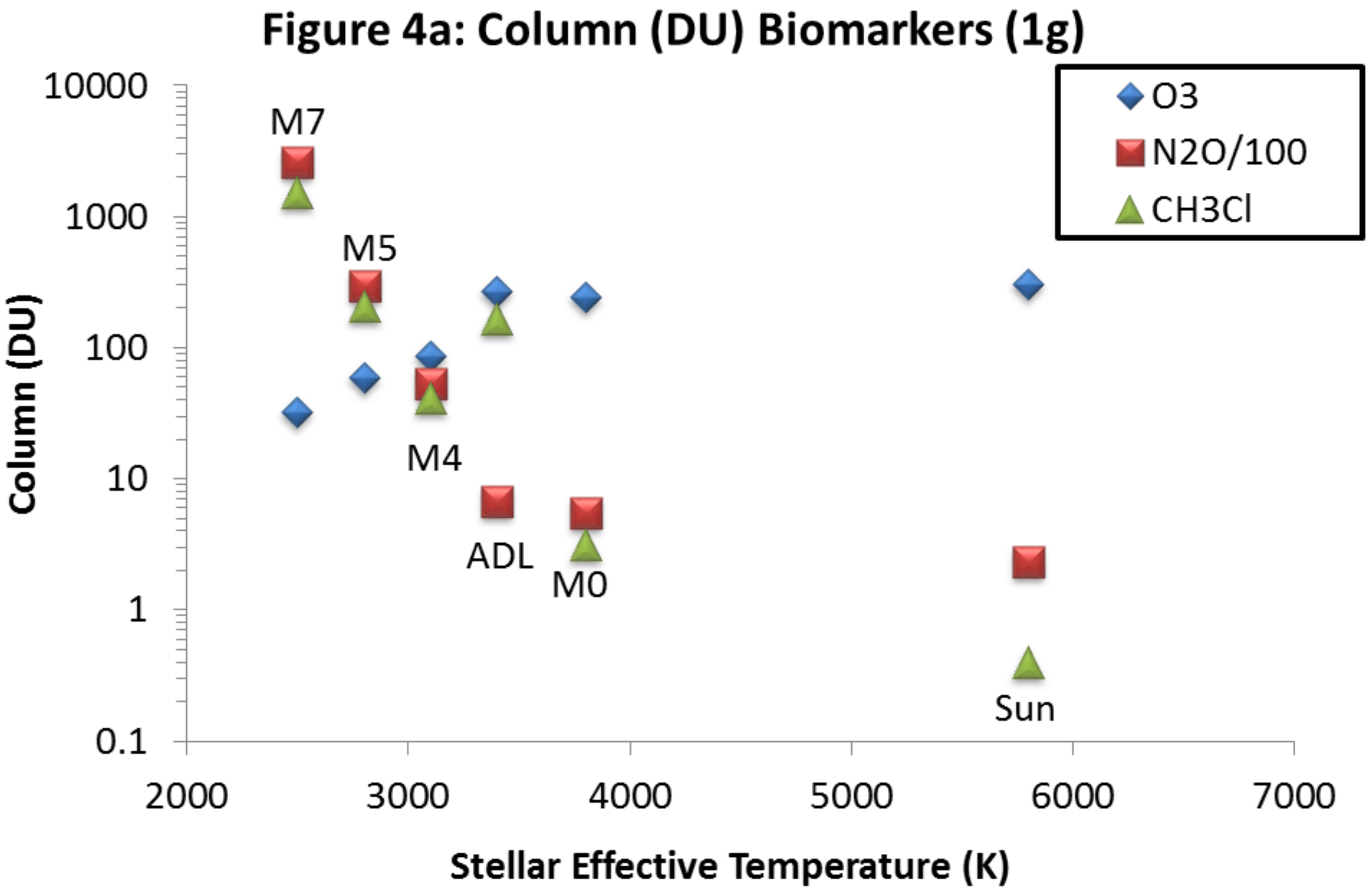

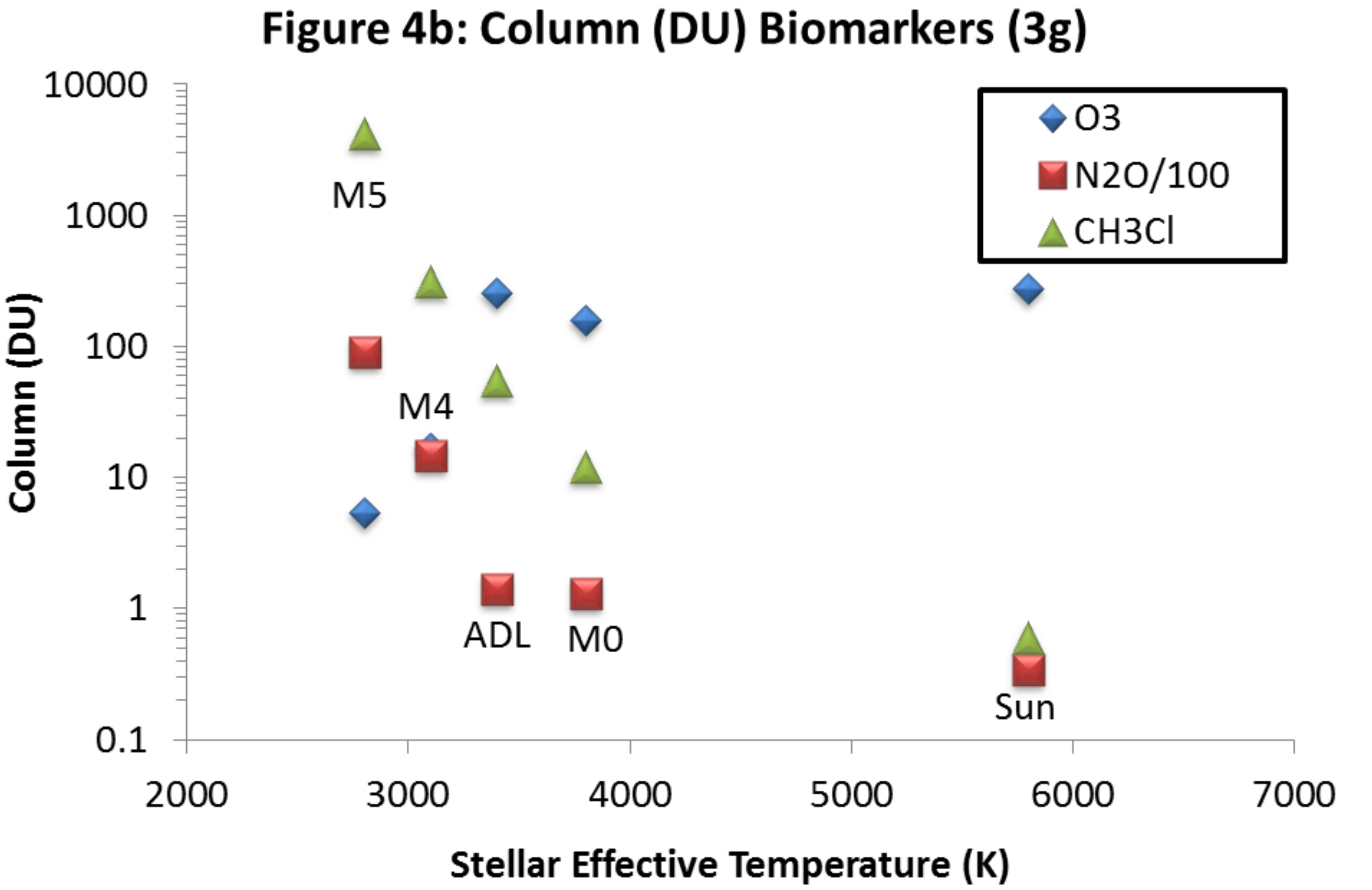

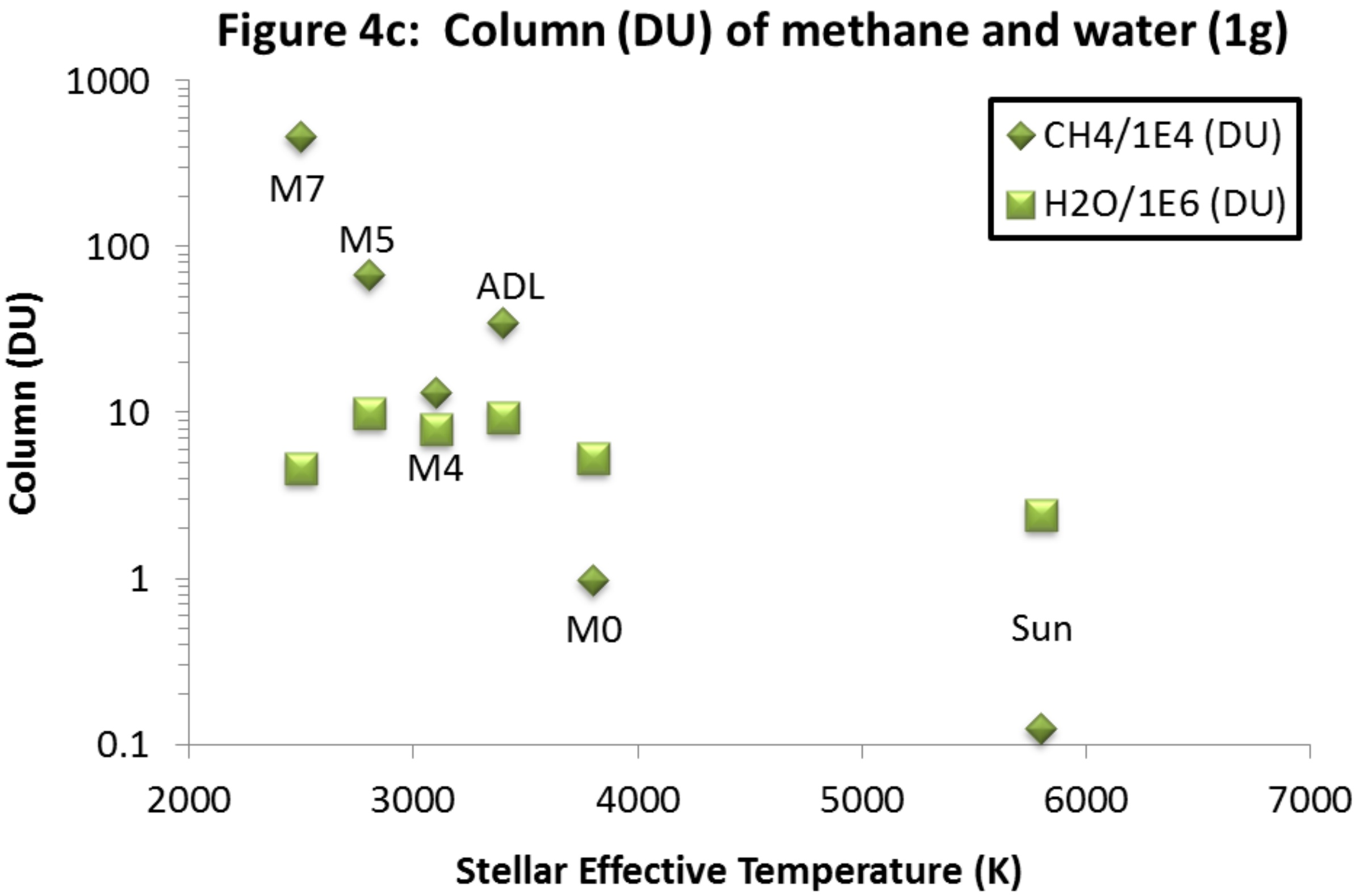
Figure 4c: Column (DU) of methane and water (1g)
Column (DU)
1000
100
10
1
0.1
2000
3000
4000
5000
6000
7000
Stellar Effective Temperature (K)
CH4/1E4 (DU)
H2O/1E6 (DU)
M7
M5
ADL
M4
M0
Sun

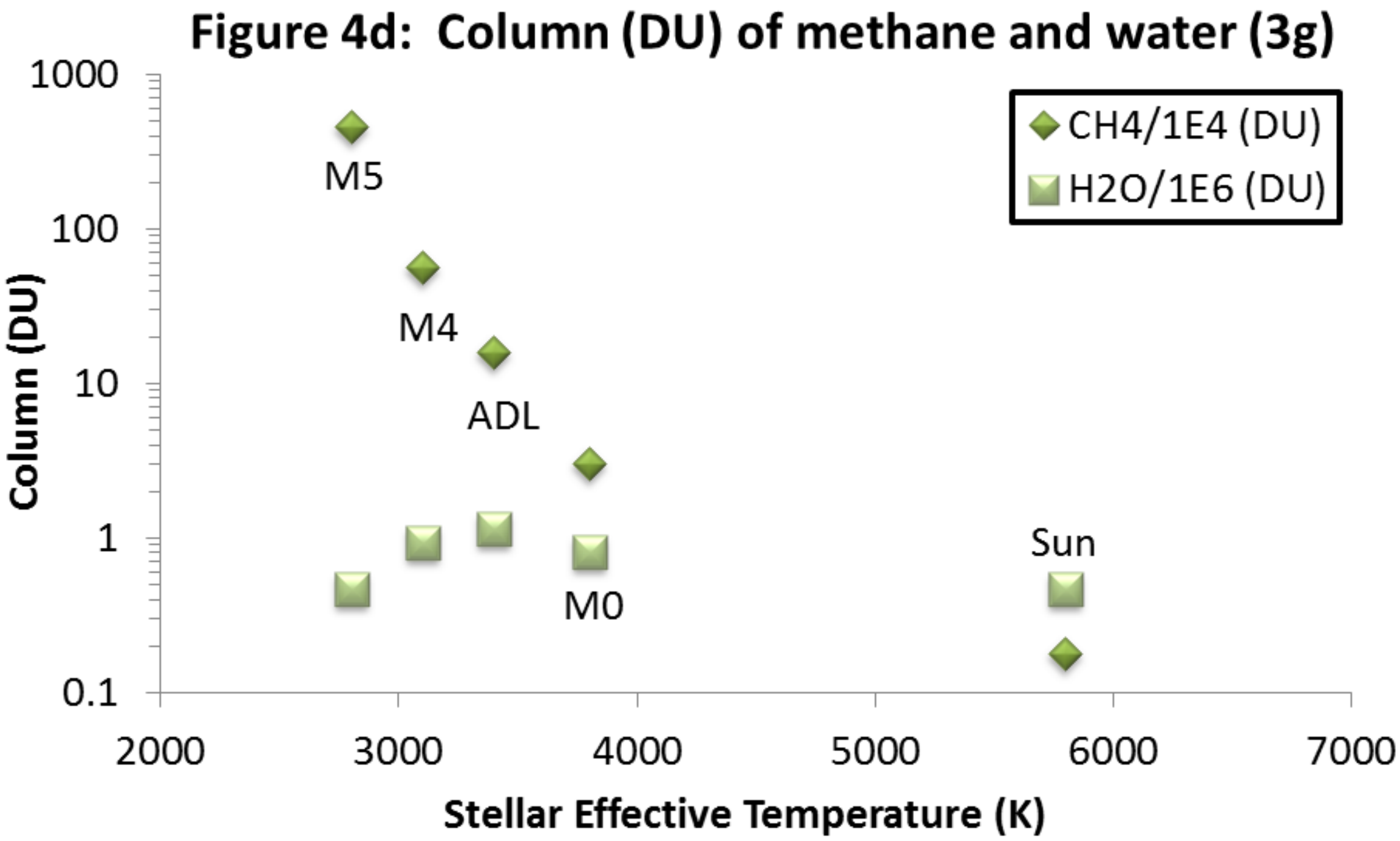
Figure 4d: Column (DU) of methane and water (3g)
Column (DU)
1000
100
10
1
0.1
2000
3000
4000
5000
6000
7000
Stellar Effective Temperature (K)
CH4/1E4 (DU)
H2O/1E6 (DU)
M5
M4
ADL
M0
Sun

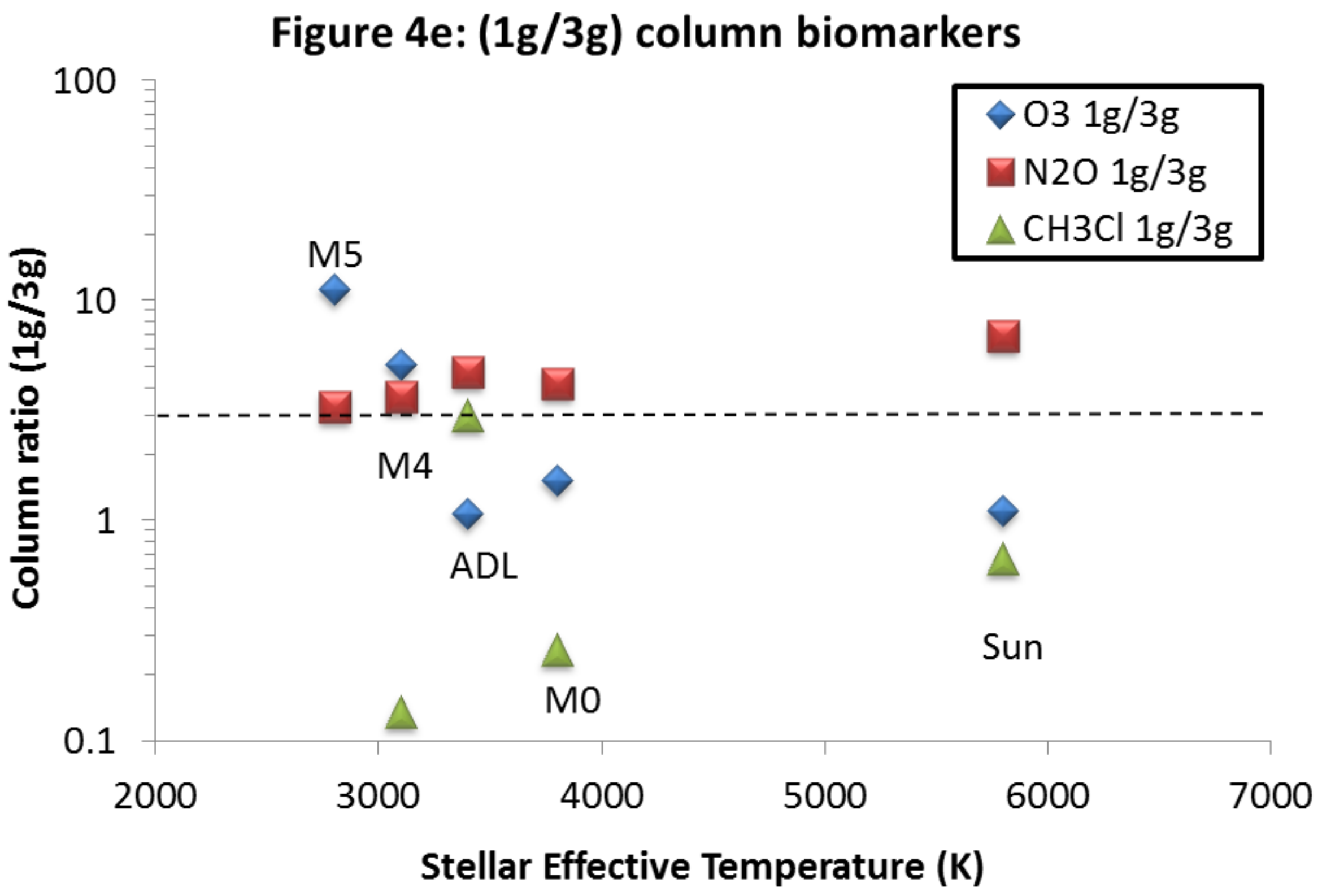

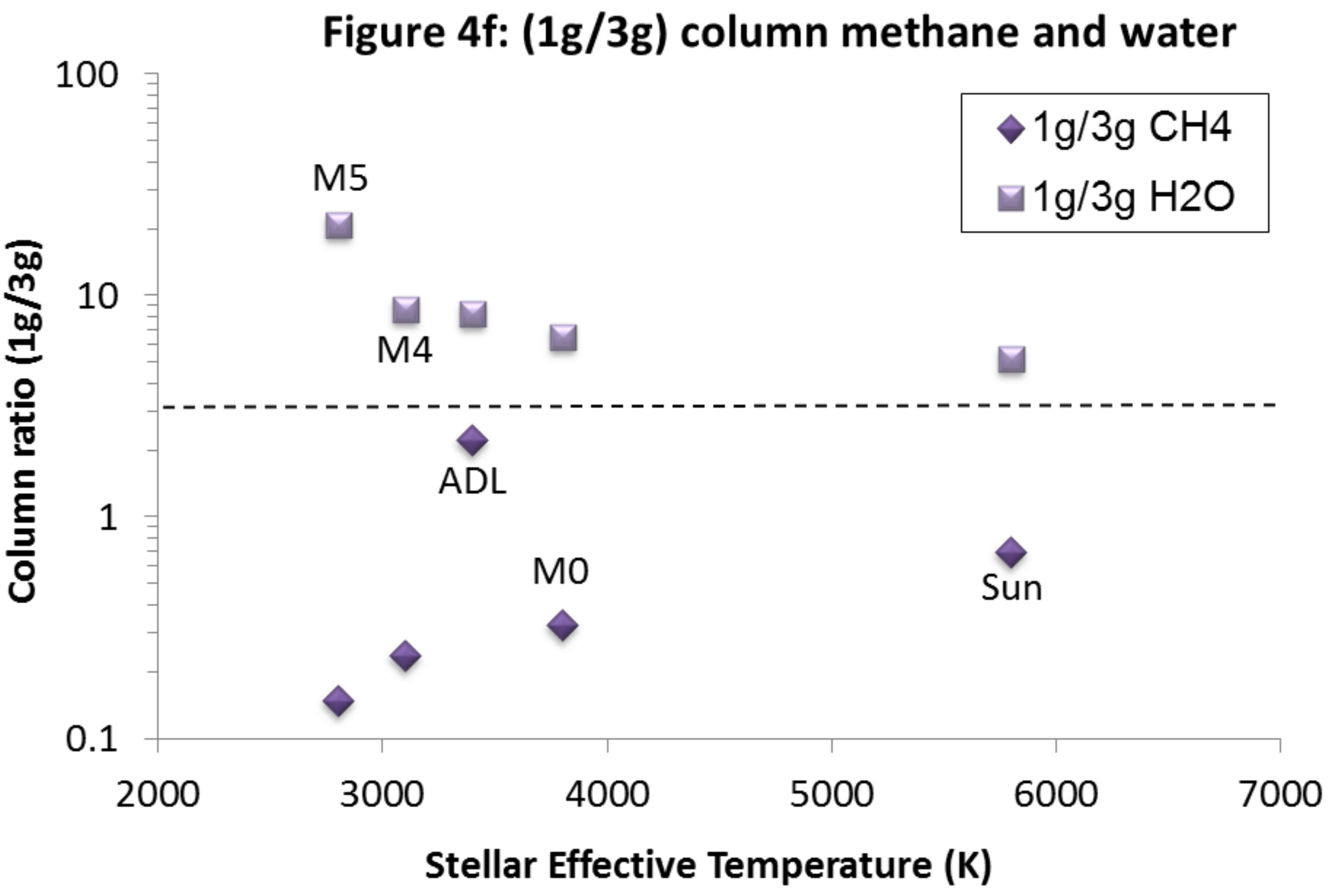

Figure 5a: Ozone Atmospheric Production and Loss (1g)

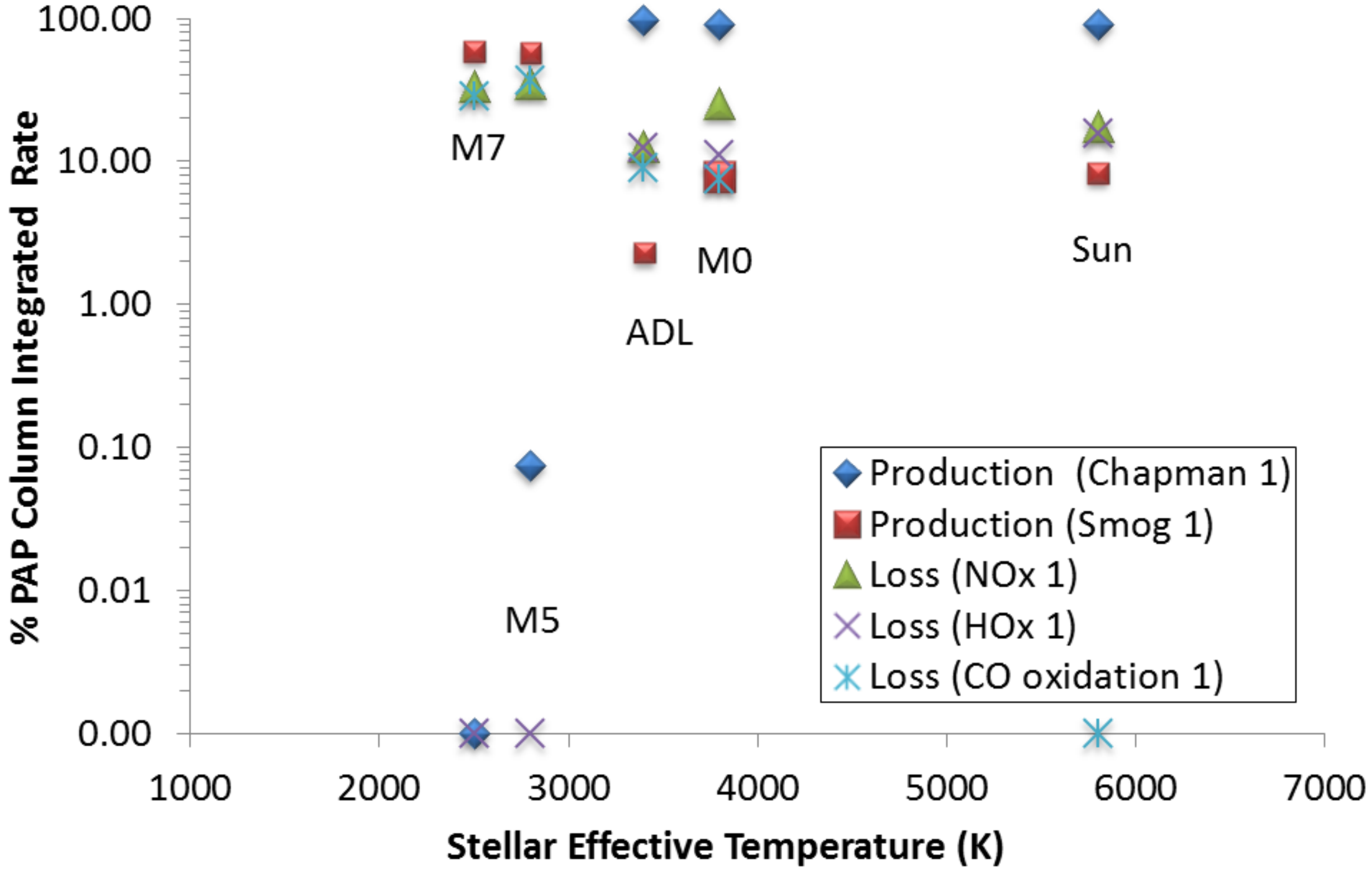

Figure 5b: Ozone Atmospheric Production and Loss (3g)

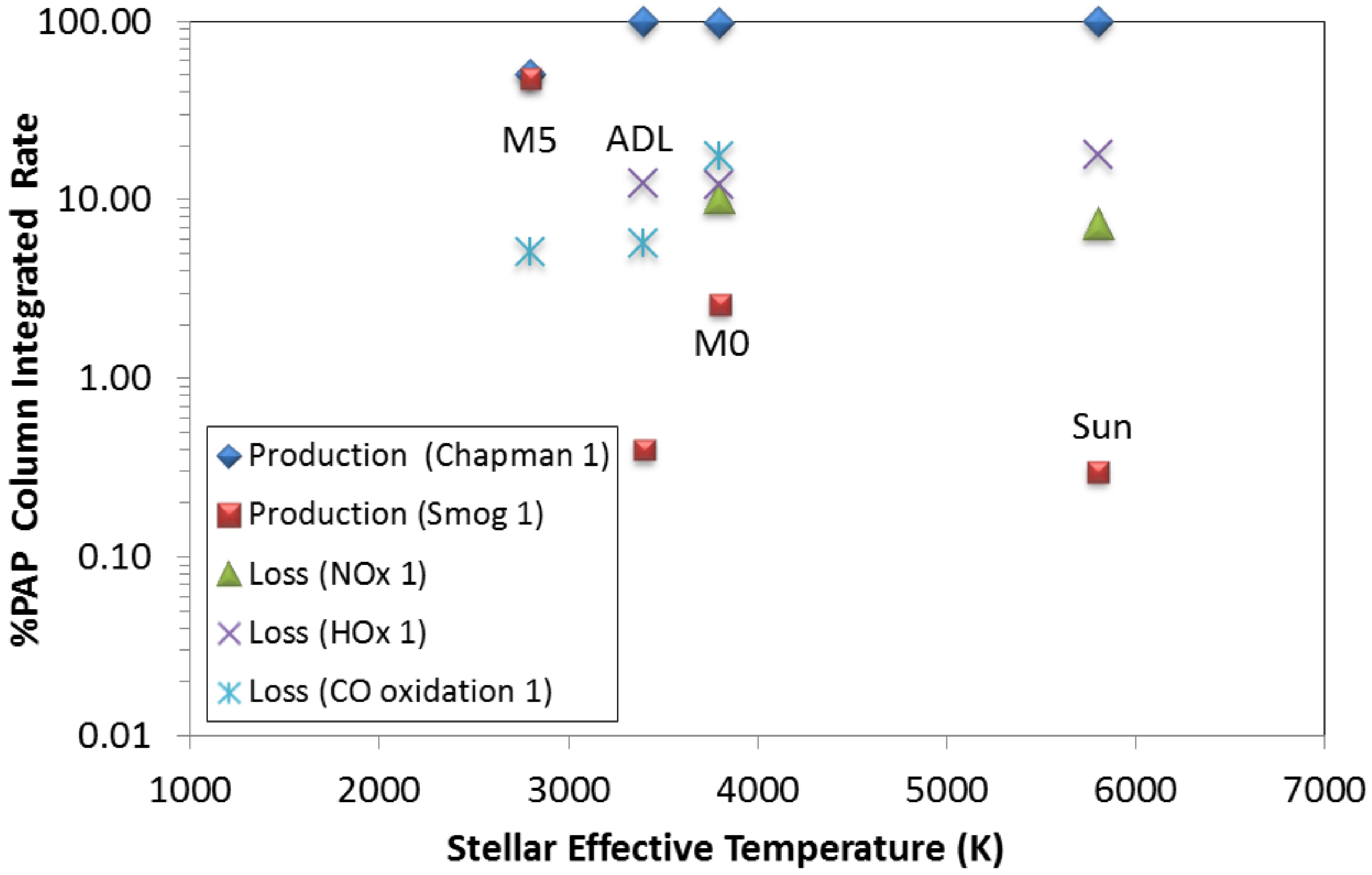

Figure 6a: 1g Sun $O_3$ Production

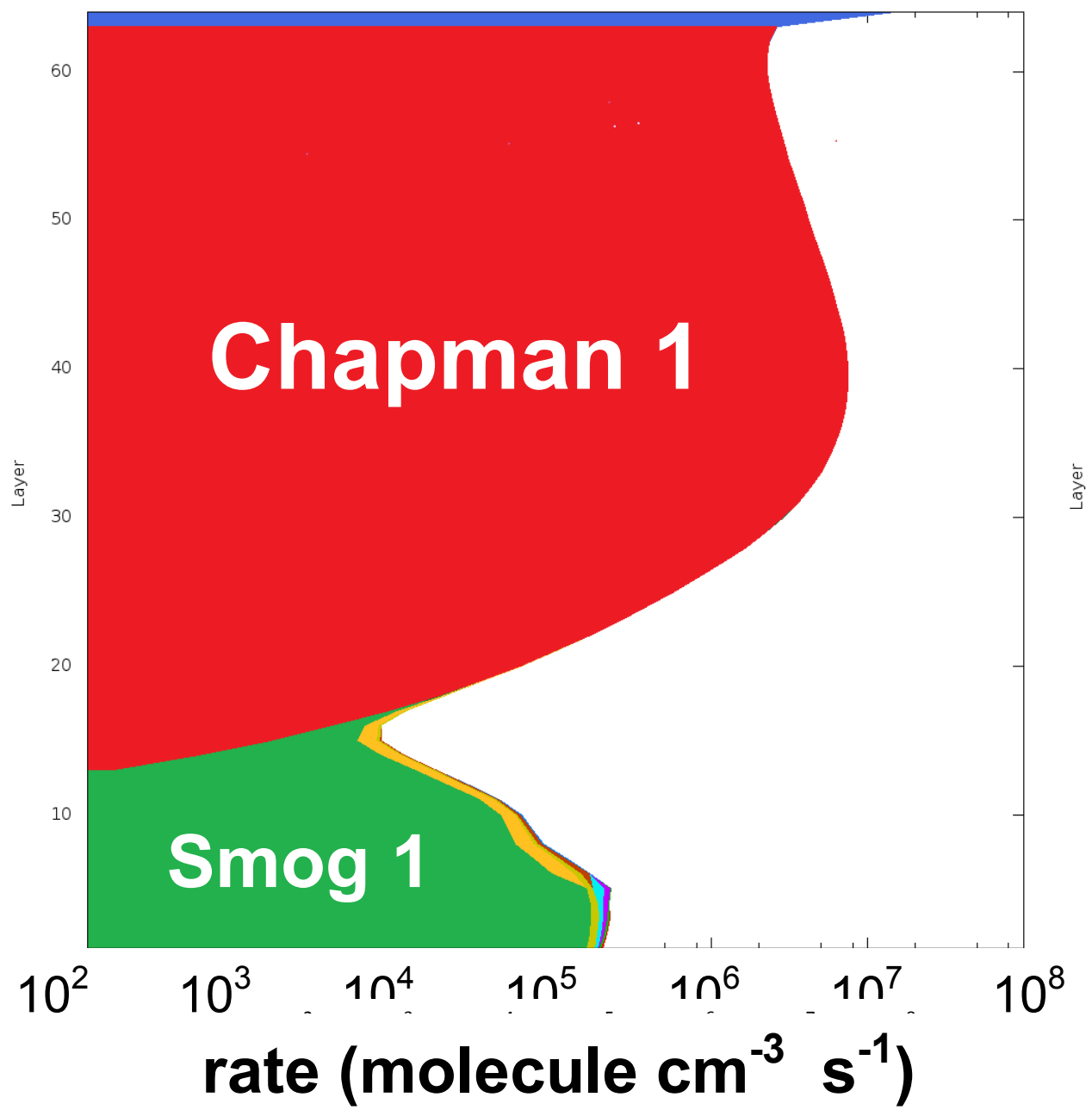

Figure 6b: 1g Sun $O_3$ Destruction

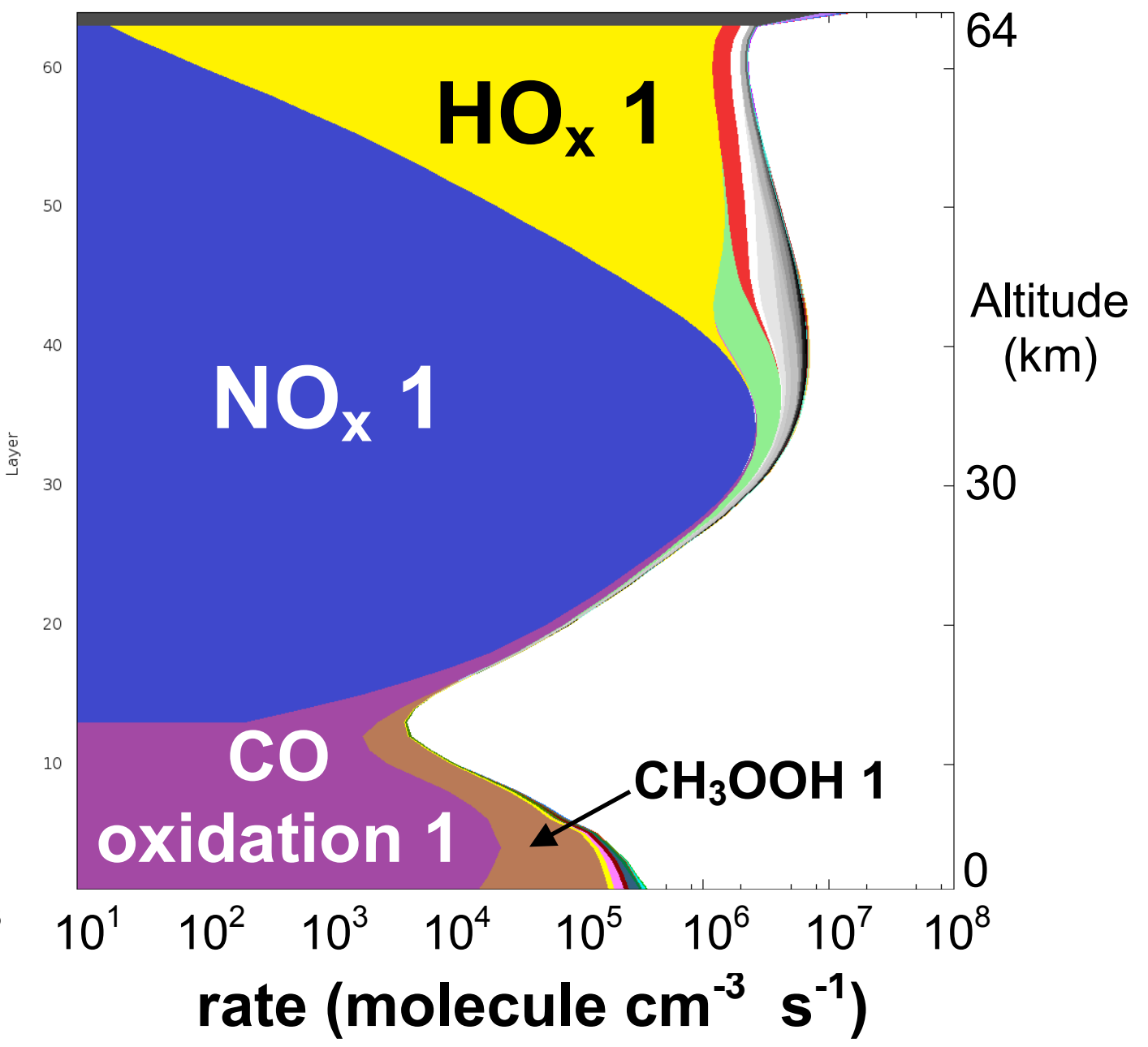

Figure 6c: 1g M7 $O_3$ Production

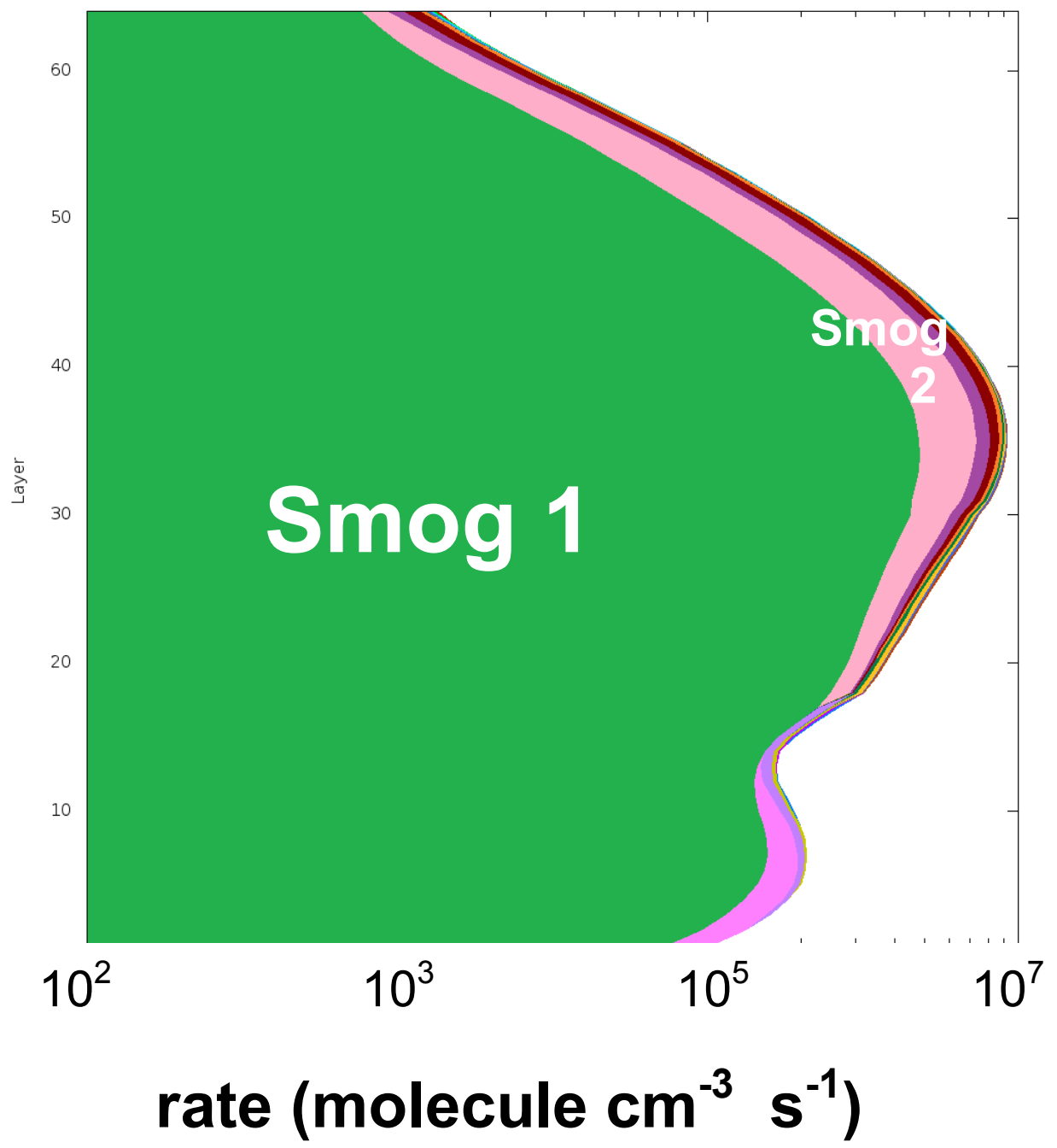

Figure 6d: 1g M7 $O_3$ Destruction

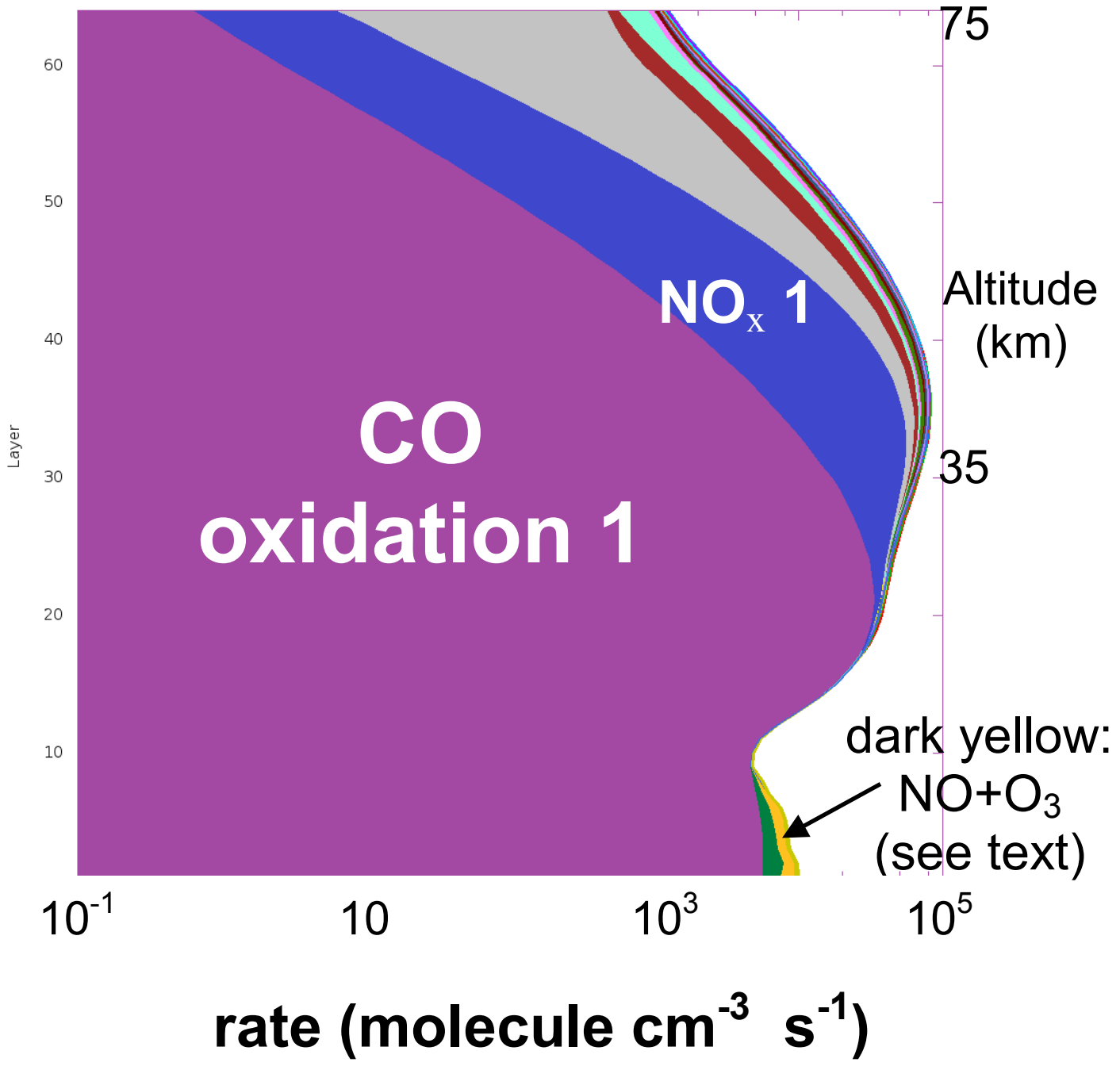

Figure 7a: 1g ADL $O_3$ Production

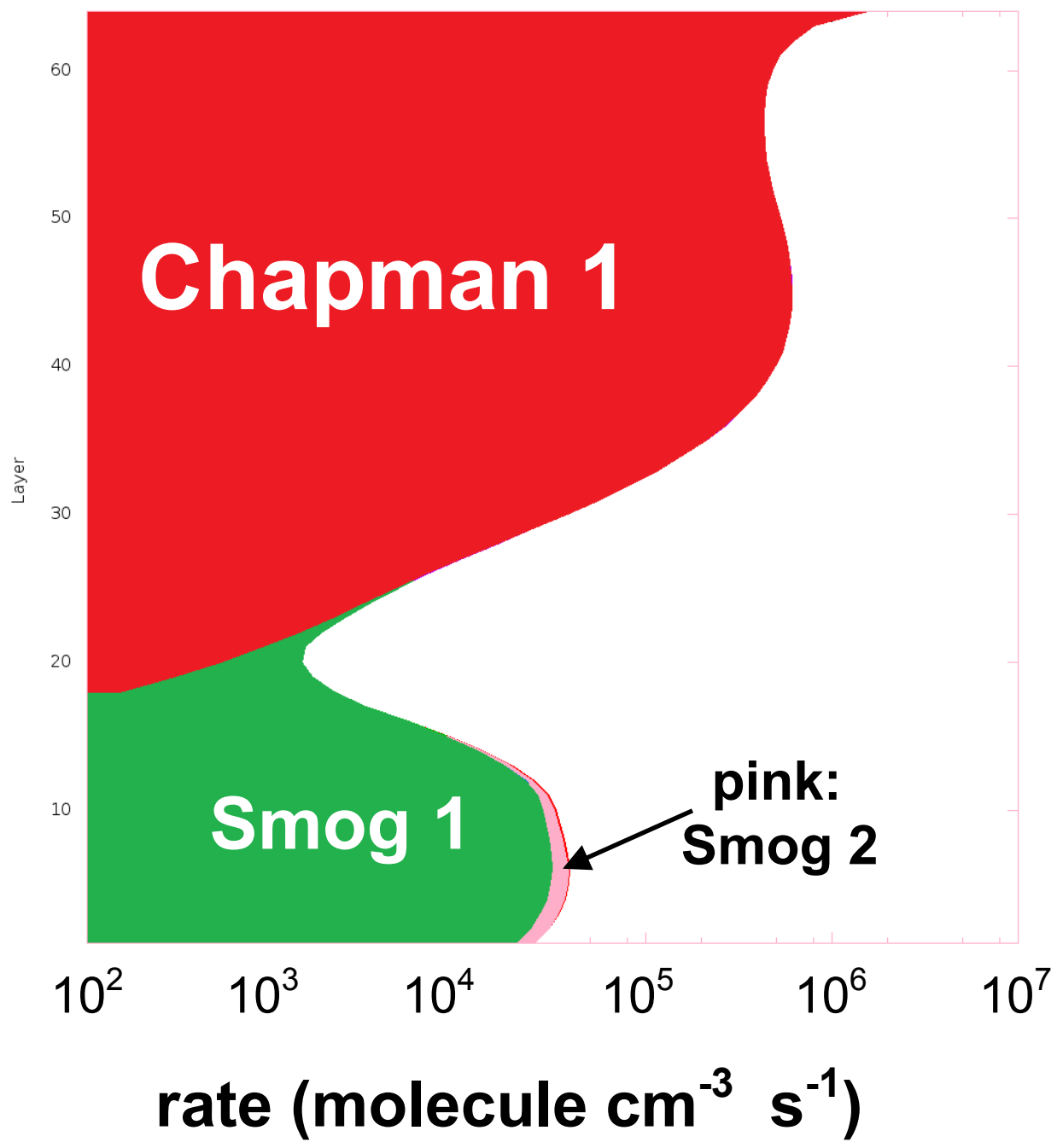

Figure 7b: 1g ADL $O_3$ Destruction

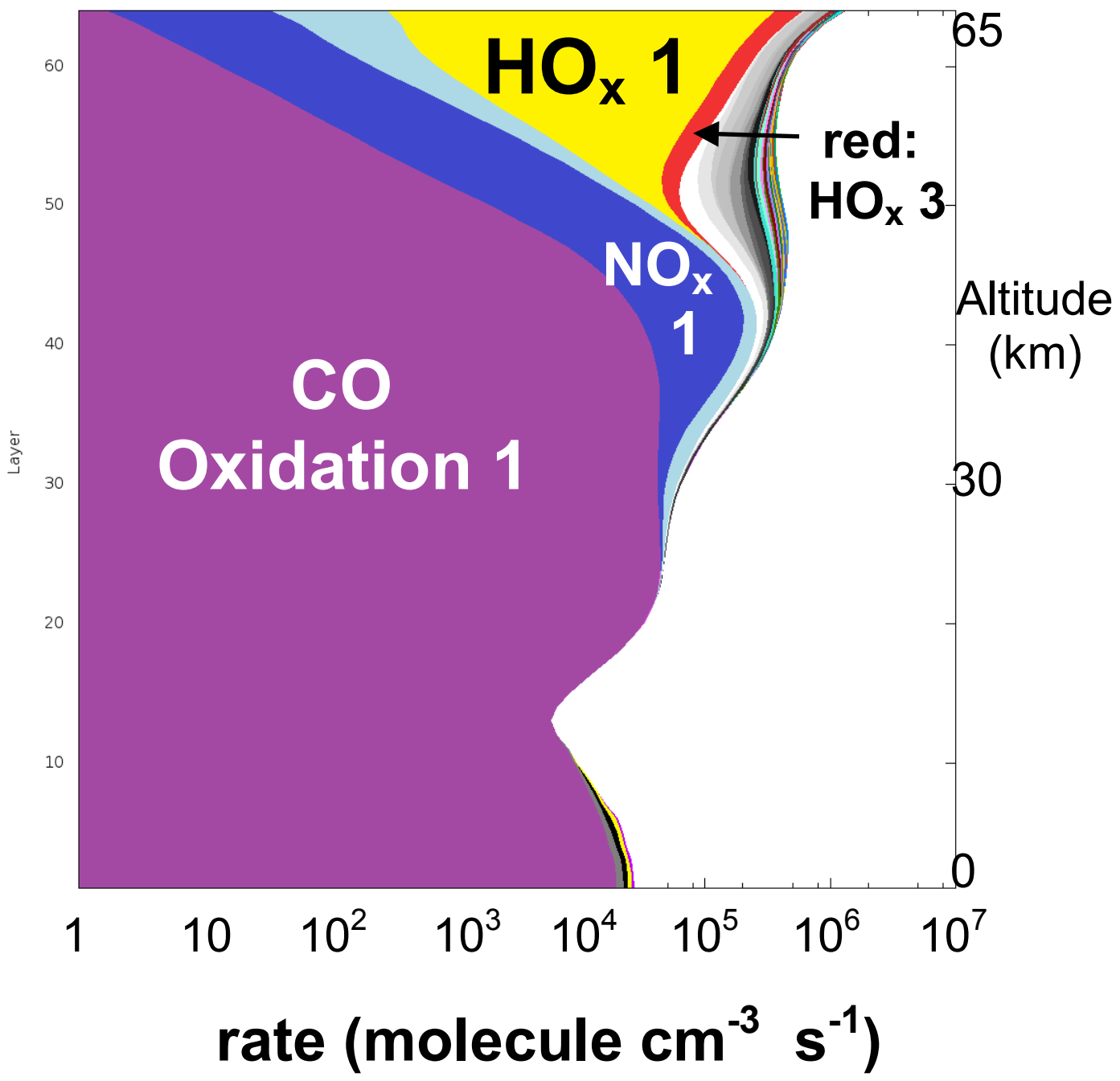

Figure 7c: 1g M5 $O_3$ Production

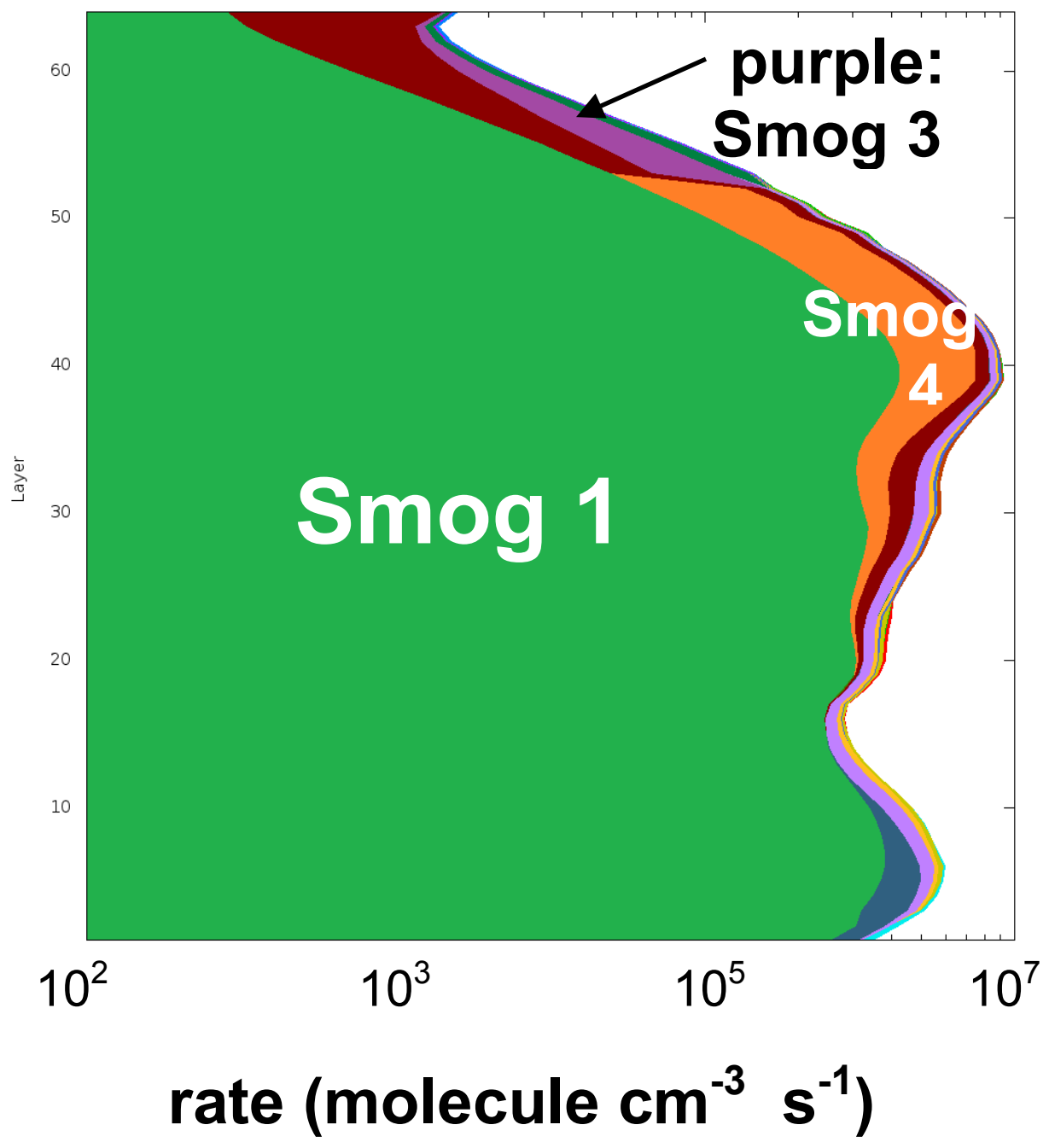

Figure 7d: 1g M5 $O_3$ Destruction

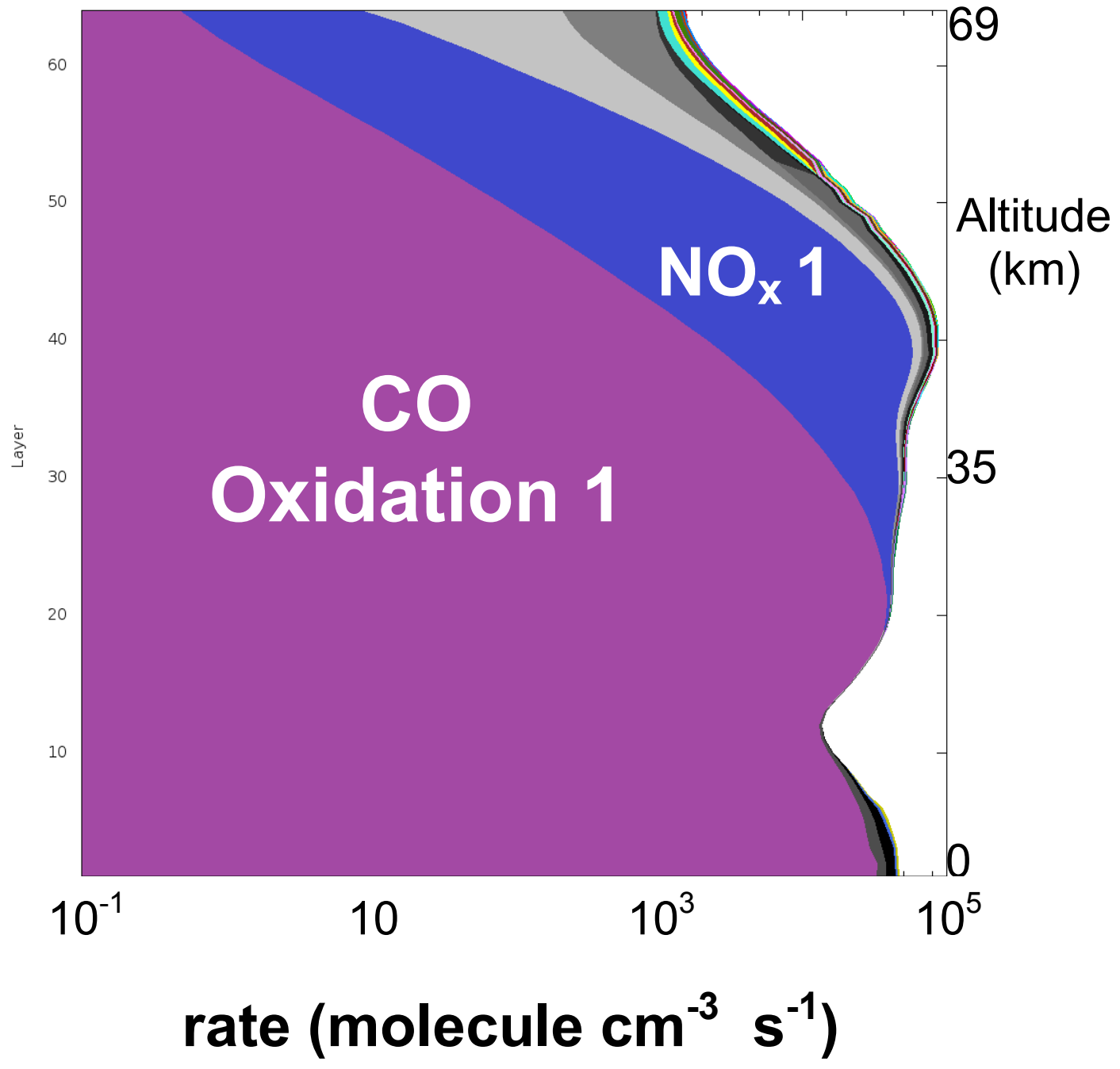

## Appendix 1: (a) Ozone Pathways

"**Found_PAP**" denotes the Column Integrated Rates (CIR) (in molecules $cm^{-2}\ s^{-1}$) of change shown for production and loss for all atmospheric pathways **found** by PAP. "**Shown_PAP**" denotes the CIR only for the pathways **shown** in this Appendix. Shown are either the 5 dominant pathways or the first pathways which together account for >90% of Found_PAP, whichever criterion is fulfilled first. "**Total _chem**" denotes the CIR as calculated in the **chemistry** scheme of the atmospheric column model. **Percent** values for a particular cycle show its individual rate as a percentage of Found_PAP.

### 1g Sun Ozone Production: Found_PAP =$1.88\times10^{13}$, Shown_PAP=$1.87\times10^{13}$, Total_chem =$1.88\times10^{13}$

| | | |
|---|---|---|
| **Chapman 1 (90.5%):**<br>$O_2 + h\nu \rightarrow O(^3P) + O(^3P)$<br>$2(O(^3P) + O_2 + M \rightarrow O_3 + M)$<br>net: $3O_2 \rightarrow 2O_3$ | **Chapman 2 (8.3%):**<br>$O(^3P) + O_2 + M \rightarrow O_3 + M$<br>net: $O(^3P) + O_2 + M \rightarrow O_3 + M$ | **Smog 1 (0.8%):**<br>$CO + OH \rightarrow CO_2 + H$<br>$H + O_2 + M \rightarrow HO_2 + M$<br>$NO + HO_2 \rightarrow NO_2 + OH$<br>$NO_2 + h\nu \rightarrow NO + O(^3P)$<br>$O(^3P) + O_2 + M \rightarrow O_3 + M$<br>net: $CO + 2O_2 \rightarrow O_3 + CO_2$ |

### 1g Sun Ozone Loss: Found_PAP =$1.81\times10^{13}$, Shown_PAP=$1.06\times10^{13}$, Total_chem=$1.88\times10^{13}$

| | | | | |
|---|---|---|---|---|
| **$NO_x$ 1 (17.6%):**<br>$O_3 + h\nu \rightarrow O_2 + O(^3P)$<br>$NO_2 + O(^3P) \rightarrow NO + O_2$<br>$NO + O_3 \rightarrow NO_2 + O_2$<br>net: $2O_3 \rightarrow 3O_2$ | **$HO_x$ 1 (15.6%):**<br>$2(O_3 + h\nu \rightarrow O_2 + O(^1D))$<br>$2(O(^1D) + N_2 \rightarrow O(^3P) + N_2)$<br>$HO_2 + O(^3P) \rightarrow OH + O_2$<br>$OH + O(^3P) \rightarrow H + O_2$<br>$H + O_2 + M \rightarrow HO_2 + M$<br>$2O_3 \rightarrow 3O_2$ | **$NO_x$ 2 (12.3%):**<br>$O_3 + h\nu \rightarrow O_2 + O(^1D)$<br>$O(^1D) + N_2 \rightarrow O(^3P) + N_2$<br>$NO_2 + O(^3P) \rightarrow NO + O_2$<br>$NO + O_3 \rightarrow NO_2 + O_2$<br>$2O_3 \rightarrow 3O_2$ | **$O_x$ 1 (6.9%):**<br>$O_3 + h\nu \rightarrow O_2 + O(^1D)$<br>$O(^1D) + N_2 \rightarrow O(^3P) + N_2$<br>$O(^3P) + O_3 \rightarrow O_2 + O_2$<br>$2O_3 \rightarrow 3O_2$ | **$HO_x$ 2 (6.3%):**<br>$2(O_3 + h\nu \rightarrow O_2 + O(^1D))$<br>$2(O(^1D) + O_2 \rightarrow O(^3P) + O_2)$<br>$HO_2 + O(^3P) \rightarrow OH + O_2$<br>$OH + O(^3P) \rightarrow H + O_2$<br>$H + O_2 + M \rightarrow HO_2 + M$<br>$2O_3 \rightarrow 3O_2$ |

### 3g Sun Ozone Production: Found_PAP = $1.28\times10^{13}$, Shown_PAP=$1.27\times10^{13}$, Total_chem = $1.28\times10^{13}$

| | |
|---|---|
| **Chapman 1**<br>(99.3%) | **Smog 1**<br>(0.3%) |

### 3g Sun Ozone Loss: Found_PAP =$1.20\times10^{13}$, Shown_PAP=$6.06\times10^{12}$, Total_chem = $1.26\times10^{13}$

| | | | | |
|---|---|---|---|---|
| **$HO_x$ 1**<br>(17.7%) | **$O_x$ 1**<br>(9.5%) | **$NO_x$ 2**<br>(8.3%) | **$O_x$ 2 (7.6%)**<br>$O_3 + h\nu \rightarrow O_2 + O(^3P)$<br>$O(^3P) + O_3 \rightarrow O_2 + O_2$<br>$2O_3 \rightarrow 3O_2$ | **$NO_x$ 1**<br>(7.3%) |

### 1g M0 Ozone Production: Found_PAP =$1.65\times10^{12}$, Shown_PAP=$1.63\times10^{12}$, Total_chem = $1.70\times10^{12}$

| | | |
|---|---|---|
| **Chapman 1 (89.2%):** | **Smog 1 (7.9%):** | **Smog 2 (1.2%):**<br>$CH_4 + OH \rightarrow CH_3 + H_2O$<br>$CH_3 + O_2 + M \rightarrow CH_3O_2 + M$<br>$CH_3O_2 + NO \rightarrow H_3CO + NO_2$<br>$H_3CO + O_2 \rightarrow H_2CO + HO_2$<br>$H_2CO + h\nu \rightarrow H_2 + CO$<br>$NO + HO_2 \rightarrow NO_2 + OH$<br>$2(NO_2 + h\nu \rightarrow NO + O(^3P))$<br>$2(O(^3P) + O_2 + M \rightarrow O_3 + M)$<br>net: $CH_4 + 4O_2 \rightarrow 2O_3 + H_2O + H_2 + CO$ |

### 1g M0 Ozone Loss: Found_PAP =$1.52\times10^{12}$, Shown_PAP=$8.92\times10^{11}$, Total_chem = $1.64\times10^{12}$

| | | | | |
|---|---|---|---|---|
| **$NO_x$ 1 (25.4%)** | **$HO_x$ 1 (11.1%)** | **$HO_x$ 3 (7.5%):**<br>$2(O_3 + h\nu \rightarrow O_2 + O(^3P))$<br>$HO_2 + O(^3P) \rightarrow OH + O_2$<br>$OH + O(^3P) \rightarrow H + O_2$<br>$H + O_2 + M \rightarrow HO_2 + M$<br>net: $2O_3 \rightarrow 3O_2$ (7.5%) | **CO-oxidation 1 (7.4%)**<br>$HO_2 + O_3 \rightarrow OH + O_2 + O_2$<br>$CO + OH \rightarrow CO_2 + H$<br>$H + O_2 + M \rightarrow HO_2 + M$<br>net: $O_3 + CO \rightarrow O_2 + CO_2$ | **$NO_x$ 1 (7.1%)** |

**3g M0 Ozone Production**: Found_PAP =$1.31 \times 10^{12}$, Shown_PAP=$1.30 \times 10^{12}$, Total_chem = $1.32 \times 10^{12}$

| Chapman 1 (96.7%) | Smog 1 (2.6%) | Chapman 2 (0.3%) |
|---|---|---|

**3g M0 Ozone Loss**: Total = Found_PAP =$9.75 \times 10^{11}$, Shown_PAP=$5.27 \times 10^{11}$, Total_chem = $1.20 \times 10^{12}$

| CO-oxidation 1 (17.6%) | $HO_x$ 1 (12.%) | $NO_x$ 1 (10.3%) | $HO_x$ 3 (8.1%) | $HO_x$ 4 (5.8%):<br>$HO_2 + O_3 \rightarrow OH + O_2 + O_2$<br>$\underline{OH + O_3 \rightarrow HO_2 + O_2}$<br>net: $2O_3 \rightarrow 3O_2$ |
|---|---|---|---|---|

**1g AD-Leo Ozone Production**: Found_PAP =$1.72 \times 10^{12}$, Shown_PAP=$1.71 \times 10^{12}$, Total_chem = $1.72 \times 10^{12}$

| Chapman 1 (97.2%) | Smog 1 (2.3%) | Smog 3 (0.3%):<br>$CH_4 + OH \rightarrow CH_3 + H_2O$<br>$CH_3 + O_2 + M \rightarrow CH_3O_2 + M$<br>$CH_3O_2 + NO \rightarrow H_3CO + NO_2$<br>$H_3CO + O_2 \rightarrow H_2CO + HO_2$<br>$2(NO_2 + h\nu \rightarrow NO + O(^3P))$<br>$NO + HO_2 \rightarrow NO_2 + OH$<br>$\underline{2(O(^3P) + O_2 + M \rightarrow O_3 + M)}$<br>net: $CH_4+4O_2 \rightarrow 2O_3+H_2CO+H_2O$ |
|---|---|---|

**1g AD-Leo Ozone Loss**: Found_PAP =$1.43 \times 10^{12}$, Shown_PAP=$6.71 \times 10^{11}$, Total_chem = $1.69 \times 10^{12}$

| $NO_x$ 1 (12.6%) | $HO_x$ 1 (12.3%) | CO-oxidation 1 (9.0%) | $HO_x$ 3 (7.5%) | $O_x$ 2 (5.4%) |
|---|---|---|---|---|

**3g AD-Leo Ozone Production:** Found_PAP =$1.56 \times 10^{12}$, Shown_PAP=$1.56 \times 10^{12}$, Total_chem = $1.56 \times 10^{12}$

| Chapman 1 (99.5%) | Smog 1 (0.4%) |
|---|---|

**3g AD-Leo Ozone Loss:** Found_PAP =$1.13 \times 10^{12}$, Shown_PAP=$4.78 \times 10^{11}$, Total_chem = $1.34 \times 10^{12}$

| $HO_x$ 1 (12.4%) | CO-oxidation 1 (10.4%) | $HO_x$ 3 (7.6%) | $HO_x$ 5 (6.2%):<br>$O_3 + h\nu \rightarrow O_2 + O(^1D)$<br>$O(^1D) + N_2 \rightarrow O(^3P) + N_2$<br>$OH + O(^3P) \rightarrow H + O_2$<br>$\underline{H + O_3 \rightarrow OH + O_2}$<br>net: $2O_3 \rightarrow 3O_2$ | CO-oxidation 2 (5.7%):<br>$O_3 + h\nu \rightarrow O_2 + O(^1D)$<br>$O(^1D) + N_2 \rightarrow O(^3P) + N_2$<br>$HO_2 + O(^3P) \rightarrow OH + O_2$<br>$CO + OH \rightarrow CO_2 + H$<br>$\underline{H + O_2 + M \rightarrow HO_2 + M}$<br>net: $O_3+CO \rightarrow O_2+CO_2$ |
|---|---|---|---|---|

**1g M5 Ozone Production**: Found_PAP =$2.92 \times 10^{11}$, Shown_PAP=$2.72 \times 10^{11}$, Total_chem = $3.24 \times 10^{11}$

| Smog 1 (57.8%) | Smog 4 (17.3%):<br>$CH_4 + OH \rightarrow CH_3 + H_2O$<br>$CH_3 + O_2 + M \rightarrow CH_3O_2 + M$<br>$CH_3O_2 + NO \rightarrow H_3CO + NO_2$<br>$H_3CO + O_2 \rightarrow H_2CO + HO_2$<br>$H_2CO + OH \rightarrow H_2O + HCO$<br>$HCO + O_2 \rightarrow HO_2 + CO$<br>$2(NO + HO_2 \rightarrow NO_2 + OH)$<br>$3(NO_2 + h\nu \rightarrow NO + O(^3P))$<br>$\underline{3(O(^3P) + O_2 + M \rightarrow O_3 + M)}$<br>$CH_4+6O_2 \rightarrow 3O_3+2H_2O+ CO$ | Chapman 1 (7.5%) | Smog 2 (6.8%) | Smog 3 (4.1%) |
|---|---|---|---|---|

**1g M5 Ozone Loss**: Found_PAP =$2.57 \times 10^{11}$, Shown_PAP=$2.19 \times 10^{11}$, Total_chem = $2.91 \times 10^{11}$

| CO-oxidation 1 (36.8%) | $NO_x$ 1 (35.2%) | Smog 5 (6.4%):<br>$O_3 + h\nu \rightarrow O_2 + O(^3P)$<br>$NO_2 + O(^3P) \rightarrow NO + O_2$<br>$NO + HO_2 \rightarrow NO_2 + OH$<br>$CO + OH \rightarrow CO_2 + H$<br>$\underline{H + O_2 + M \rightarrow HO_2 + M}$<br>$O_3 + CO \rightarrow O_2 + CO_2$ | Smog 6 (3.8%):<br>$3(O_3 + h\nu \rightarrow O_2 + O(^3P))$<br>$3(NO_2 + O(^3P) \rightarrow NO + O_2)$<br>$2(NO + HO_2 \rightarrow NO_2 + OH)$<br>$CH_4 + OH \rightarrow CH_3 + H_2O$<br>$CH_3 + O_2 + M \rightarrow CH_3O_2 + M$<br>$CH_3O_2 + NO \rightarrow H_3CO + NO_2$<br>$H_3CO + O_2 \rightarrow H_2CO + HO_2$<br>$H_2CO + OH \rightarrow H_2O + HCO$<br>$\underline{HCO + O_2 \rightarrow HO_2 + CO}$<br>$3O_3+CH_4 \rightarrow 2H_2O+CO+3O_2$ | $NO_x$ 2 (3.0%) |
|---|---|---|---|---|

**3g M5 Ozone Production**: Found_PAP =$1.35 \times 10^{11}$, Shown_PAP=$1.32 \times 10^{11}$, Total_chem = $1.42 \times 10^{11}$

| **Chapman 1 (47.8%)** | **Smog 1 (47.6%)** | **Chapman 2 (2.4%)** |
|---|---|---|

**3g M5 Ozone Loss**: Found_PAP =$1.19 \times 10^{11}$, Shown_PAP=$1.08 \times 10^{11}$, Total_chem = $1.28 \times 10^{11}$

| | | | |
|---|---|---|---|
| **$CH_3OOH$ 1 (46.8%):**<br>$O_3 + h\nu \rightarrow O_2 + O(^1D)$<br>$CH_4 + O(^1D) \rightarrow CH_3 + OH$<br>$CH_3 + O_2 + M \rightarrow CH_3O_2 + M$<br>$CO + OH \rightarrow CO_2 + H$<br>$H + O_2 + M \rightarrow HO_2 + M$<br>$\underline{CH_3O_2 + HO_2 \rightarrow CH_3OOH + O_2}$<br>net: $O_3 + CH_4 + CO \rightarrow CH_3OOH + CO_2$ | **$CH_4$-$O_3$ oxidation (6.2%):**<br>$O_3 + h\nu \rightarrow O_2 + O(^1D)$<br>$\underline{CH_4 + O(^1D) \rightarrow H_2CO + H_2}$<br>net: $O_3 + CH_4 \rightarrow H_2CO + H_2 + O_2$ | **CO-oxidation 3 (5.1%):**<br>$O_3 + h\nu \rightarrow O_2 + O(^3P)$<br>$HO_2 + O(^3P) \rightarrow OH + O_2$<br>$CO + OH \rightarrow CO_2 + H$<br>$\underline{H + O_2 + M \rightarrow HO_2 + M}$<br>$O_3 + CO \rightarrow O_2 + CO_2$ | **$H_2O_2$ cycle (4.6%):**<br>$O_3 + h\nu \rightarrow O_2 + O(^1D)$<br>$H_2O + O(^1D) \rightarrow OH + OH$<br>$2(CO + OH \rightarrow CO_2 + H)$<br>$2(H + O_2 + M \rightarrow HO_2 + M)$<br>$\underline{HO_2 + HO_2 \rightarrow H_2O_2 + O_2}$<br>net: $O_3 + H_2O + 2CO \rightarrow H_2O_2 + 2CO_2$ |

**1g M7 Ozone Production**: Found_PAP =$2.79 \times 10^{11}$, Shown_PAP=$2.59 \times 10^{11}$, Total_chem = $2.92 \times 10^{11}$

| | | | |
|---|---|---|---|
| **Smog 1 (58.8%)** | **Smog 3 (22.9%)** | **$CH_3OOH$ 2 smog cycle (6.0%);**<br>$2(NO_2 + h\nu \rightarrow NO + O(^3P))$<br>$2(O(^3P) + O_2 + M \rightarrow O_3 + M)$<br>$NO + HO_2 \rightarrow NO_2 + OH$<br>$CH_3OOH + OH \rightarrow CH_3O_2 + H_2O$<br>$CH_3O_2 + NO \rightarrow H_3CO + NO_2$<br>$\underline{H_3CO + O_2 \rightarrow H_2CO + HO_2}$<br>net: $CH_3OOH + 3O_2 \rightarrow 2O_3 + H_2CO + H_2O$ | **HCHO cycle (4.9%)**<br>$NO_2 + h\nu \rightarrow NO + O(^3P)$<br>$NO + HO_2 \rightarrow NO_2 + OH$<br>$H_2CO + OH \rightarrow H_2O + HCO$<br>$HCO + O_2 \rightarrow HO_2 + CO$<br>$\underline{O(^3P) + O_2 + M \rightarrow O_3 + M}$<br>net: $H_2CO + 2O_2 \rightarrow O_3 + H_2O + CO$ |

**1g M7 Ozone Loss**: Found_PAP =$2.43 \times 10^{11}$, Shown_PAP=$2.01 \times 10^{11}$, Total_chem = $2.74 \times 10^{11}$

| | | | | |
|---|---|---|---|---|
| **$NO_x$ 1 (33.1%)** | **CO oxidation 1 (28.3%)** | **Smog 5 (11.5%)** | **Smog 7 (6.4%):**<br>$2(O_3 + h\nu \rightarrow O_2 + O(^3P))$<br>$2(NO_2 + O(^3P) \rightarrow NO + O_2)$<br>$NO + HO_2 \rightarrow NO_2 + OH$<br>$CH_4 + OH \rightarrow CH_3 + H_2O$<br>$CH_3 + O_2 + M \rightarrow CH_3O_2 + M$<br>$CH_3O_2 + NO \rightarrow H_3CO + NO_2$<br>$\underline{H_3CO + O_2 \rightarrow H_2CO + HO_2}$<br>net: $2O_3 + CH_4 \rightarrow H_2CO + H_2O + 2O_2$ | **$NO_x$ 3 (3.3%):**<br>$2(NO + O_3 \rightarrow NO_2 + O_2)$<br>$NO_2 + h\nu \rightarrow NO + O(^3P)$<br>$\underline{NO_2 + O(^3P) \rightarrow NO + O_2}$<br>net: $2O_3 \rightarrow 3O_2$ |

## 1b. Nitrous Oxide Pathways (Sun only)

**1g Sun Nitrous Oxide Loss**: Found_PAP =$9.63 \times 10^{8}$, Shown_PAP=$8.78 \times 10^{8}$, Total_chem = $1.15 \times 10^{9}$

| | | | | |
|---|---|---|---|---|
| **$N_2O$-$NO_x$ (69.5%):**<br>$2(N_2O + h\nu \rightarrow N_2 + O(^3P))$<br>$O(^3P) + O_2 + M \rightarrow O_3 + M$<br>$NO_2 + O(^3P) \rightarrow NO + O_2$<br>$\underline{NO + O_3 \rightarrow NO_2 + O_2}$<br>net: $2N_2O \rightarrow O_2 + 2N_2$ | **$N_2O$-$O_x$ (9.6%):**<br>$2(N_2O + h\nu \rightarrow N_2 + O(^3P))$<br>$O(^3P) + O_2 + M \rightarrow O_3 + M$<br>$\underline{O(^3P) + O_3 \rightarrow O_2 + O_2}$<br>net: $2N_2O \rightarrow O_2 + 2N_2$ | **$N_2$-$O(^1D)$ (5.4%):**<br>$2(N_2O + O(^1D) \rightarrow N_2 + O_2)$<br>$2(O(^3P) + O_2 + M \rightarrow O_3 + M)$<br>$2(O_3 + h\nu \rightarrow O_2 + O(^1D))$<br>$\underline{O_2 + h\nu \rightarrow O(^3P) + O(^3P)}$<br>net: $2N_2O \rightarrow O_2 + 2N_2$ | **$N_2O$-$HO_x$-1 (3.4%):**<br>$2(N_2O + h\nu \rightarrow N_2 + O(^3P))$<br>$2(O(^3P) + O_2 + M \rightarrow O_2 + M)$<br>$OH + O_3 \rightarrow HO_2 + O_2$<br>$\underline{HO_2 + O_3 \rightarrow OH + O_2 + O_2}$<br>net: $2N_2O \rightarrow O_2 + 2N_2$ | **$N_2O$-$HO_x$-2 (3.3%):**<br>$2(N_2O + h\nu \rightarrow N_2 + O(^3P)$<br>$O(^3P) + O_2 + M \rightarrow O_3 + M$<br>$HO_2 + O(^3P) \rightarrow OH + O_2$<br>$\underline{OH + O_3 \rightarrow HO_2 + O_2}$<br>net: $2N_2O \rightarrow O_2 + 2N_2$ |

## 1c. Methane Pathways

**1g Sun Methane Loss**: Found_PAP =1.15x10$^{10}$, Shown_PAP=7.68x10$^{9}$, Total_chem = 1.24x10$^{11}$

| **Oxidation $3O_2$-a (29.6%):** | **Oxidation $6O_2$ (23.0%)** | **Oxidation $2O_2$-a (6.2%)** | **Oxidation $8O_2$ (6.0%)** | **Oxidation $3O_2$-b (2.2%)** |
|---|---|---|---|---|
| $2(CH_4+OH \rightarrow CH_3+H_2O)$<br>$2(CH_3+O_2+M \rightarrow CH_3O_2+M)$<br>$2(CH_3O_2+NO \rightarrow H_3CO+NO_2)$<br>$2(H_3CO+O_2 \rightarrow H_2CO+HO_2)$<br>$2(H_2CO + h\nu \rightarrow H_2 + CO)$<br>$NO + HO_2 \rightarrow NO_2 + OH$<br>$3(NO_2 + h\nu \rightarrow NO + O(^3P))$<br>$3(O(^3P) + O_2 + M \rightarrow O_3 + M)$<br>$3(HO_2+O_3 \rightarrow OH+O_2+O_2)$<br>$2(CO + OH \rightarrow CO_2 + H)$<br>$2(H + O_2 + M \rightarrow HO_2 +M)$<br>net: $2CH_4+3O_2 \rightarrow 2H_2O+ 2H_2+2CO_2$ | $CH_4 + OH \rightarrow CH_3 + H2O$<br>$CH_3 + O_2 + M \rightarrow CH_3O_2 + M$<br>$CH_3O_2 + NO \rightarrow H_3CO + NO_2$<br>$H_3CO + O_2 \rightarrow H_2CO + HO_2$<br>$H_2CO + h\nu \rightarrow H_2 + CO$<br>$CO + OH \rightarrow CO_2 + H$<br>$H + O_2 + M \rightarrow HO_2 +M$<br>$2(NO + HO_2 \rightarrow NO_2 + OH)$<br>$3(NO_2 + h\nu \rightarrow NO + O(^3P))$<br>$3(O(^3P) + O_2 + M \rightarrow O_3 + M)$<br>net: $CH_4 + 6O_2 \rightarrow H_2O + 3O_3 + H_2 + CO_2$ | $CH_4 + OH \rightarrow CH_3 + H2O$<br>$CH_3 + O_2 + M \rightarrow CH_3O_2 + M$<br>$CH_3O_2 + NO \rightarrow H_3CO + NO_2$<br>$H_3CO + O_2 \rightarrow H_2CO + HO_2$<br>$H_2CO + OH \rightarrow H_2O + HCO$<br>$NO + HO_2 \rightarrow NO_2 + OH$<br>$2(NO_2 + h\nu \rightarrow NO + O(^3P))$<br>$2(O(^3P) + O_2 + M \rightarrow O_3 + M)$<br>$HCO + O_2 \rightarrow HO_2 + CO$<br>$2(HO_2+O_3 \rightarrow OH+O_2+O_2)$<br>$CO + OH \rightarrow CO_2 + H$<br>$H + O_2 + M \rightarrow HO_2 +M$<br>net: $CH_4 +2O_2 \rightarrow 2H_2O + CO_2$ | $CH_4 + OH \rightarrow CH_3 + H_2O$<br>$CH_3 + O_2 + M \rightarrow CH_3O_2 + M$<br>$CH_3O_2 + NO \rightarrow H_3CO + NO_2$<br>$H_3CO + O_2 \rightarrow H_2CO + HO_2$<br>$H_2CO + OH \rightarrow H_2O + HCO$<br>$HCO + O_2 \rightarrow HO_2 + CO$<br>$CO + OH \rightarrow CO_2 + H$<br>$H + O_2 + M \rightarrow HO_2 +M$<br>$3(NO + HO_2 \rightarrow NO_2 + OH)$<br>$4(NO_2 + h\nu \rightarrow NO + O(^3P))$<br>$4(O(^3P) + O_2 + M \rightarrow O_3 + M)$<br>net: $CH_4 + 8O_2 \rightarrow 2H_2O + 4O_3 + CO_2$ | $2(CH_4 + OH \rightarrow CH_3 + H_2O)$<br>$2(CH_3+O_2+M \rightarrow CH_3O_2+M)$<br>$2(CH_3O_2+NO \rightarrow H_3CO+NO_2)$<br>$2(H_3CO+O_2 \rightarrow H_2CO+HO_2)$<br>$2(H_2CO + h\nu \rightarrow H_2 + CO)$<br>$2(CO + OH \rightarrow CO_2 + H)$<br>$2(H + O_2 + M \rightarrow HO_2 +M)$<br>$4(NO + HO_2 \rightarrow NO_2 + OH)$<br>$3(NO_2 + h\nu \rightarrow NO + O(^3P))$<br>$3(NO_2 + O(^3P) \rightarrow NO + O_2)$<br>net: $2CH_4 + 3O_2 \rightarrow 2H_2O + 2H_2 + 2CO_2$ |

**3g Sun Methane Loss**: Found_PAP =3.21x10$^{10}$, Shown_PAP=1.63x10$^{10}$, Total_chem = 1.24x10$^{11}$

| **Oxidation $CH_3OOH$-a (26.5%)** | **Oxidation $2O_2$-b (7.8%)** | **Oxidation $O_3$-a (5.8%)** | **Oxidation $2O_2$-c (5.4%)** | **Oxidation $O_3$-b (5.3%)** |
|---|---|---|---|---|
| $CH_4 + OH \rightarrow CH_3 + H_2O$<br>$CH_3 + O_2 + M \rightarrow CH_3O_2 + M$<br>$CH_3O_2+HO_2 \rightarrow CH_3OOH+O_2$<br>$H_2O + O(^1D) \rightarrow OH + OH$<br>$CO + OH \rightarrow CO_2 + H$<br>$H + O_2 + M \rightarrow HO_2 +M$<br>$O_3 + h\nu \rightarrow O_2 + O(^1D)$<br>net: $CH_4 + O_3 + CO \rightarrow CH_3OOH + CO_2$ | $CH_4 + OH \rightarrow CH_3 + H_2O$<br>$CH_3 + O_2 + M \rightarrow CH_3O_2 + M$<br>$CH_3O_2+OH \rightarrow H_3CO+HO_2$<br>$H_3CO + O_2 \rightarrow H_2CO + HO_2$<br>$H_2CO + OH \rightarrow H_2O + HCO$<br>$HCO + O_2 \rightarrow HO_2 + CO$<br>$CO + OH \rightarrow CO_2 + H$<br>$H + O_2 + M \rightarrow HO_2 +M$<br>$4(HO_2 + O(^3P) \rightarrow OH + O_2)$<br>$2(O_2 + h\nu \rightarrow O(^3P) + O(^3P))$<br>net: $CH_4 +2O_2 \rightarrow 2H_2O+CO_2$ | $CH_4 + OH \rightarrow CH_3 + H_2O$<br>$CH_3 + O_2 + M \rightarrow CH_3O_2 + M$<br>$CH_3O_2 + NO \rightarrow H_3CO + NO_2$<br>$H_3CO + O_2 \rightarrow H_2CO + HO_2$<br>$NO_2 + h\nu \rightarrow NO + O(^3P)$<br>$O(^3P) + O_2 + M \rightarrow O_3 + M$<br>$H_2CO + h\nu \rightarrow H_2 + CO$<br>$CO + OH \rightarrow CO_2 + H$<br>$H + O_2 + M \rightarrow HO_2 +M$<br>$2(HO_2+O_3 \rightarrow OH+O_2+O_2)$<br>net: $CH_4 +O_3 \rightarrow H_2O+H_2+CO_2$ | $CH_4 + O(^1D) \rightarrow CH_3 + OH$<br>$CH_3 + O_2 + M \rightarrow CH_3O_2 + M$<br>$CO + OH \rightarrow CO_2 + H$<br>$H + O_2 + M \rightarrow HO_2 +M$<br>$CH_3O_2+OH \rightarrow H_3CO+HO_2$<br>$H_3CO + O_2 \rightarrow H_2CO + HO_2$<br>$3(HO_2 + O(^3P) \rightarrow OH + O_2)$<br>$OH + HO_2 \rightarrow H_2O + O_2$<br>$H_2CO + OH \rightarrow H_2O + HCO$<br>$HCO + O_2 \rightarrow HO_2 + CO$<br>$O(^3P) + O_2 + M \rightarrow O_3 + M$<br>$O_3 + h\nu \rightarrow O_2 + O(^1D)$<br>$2(O_2 + h\nu \rightarrow O(^3P) + O(^3P))$<br>net: $CH_4 +2O_2 \rightarrow 2H_2O+CO_2$ | $CH_4 + OH \rightarrow CH_3 + H_2O$<br>$CH_3 + O_2 + M \rightarrow CH_3O_2 + M$<br>$CH_3O_2+HO_2 \rightarrow CH_3OOH+O_2$<br>$CH_3OOH+h\nu \rightarrow H_3CO+OH$<br>$H_3CO+O_2 \rightarrow H_2CO+HO_2$<br>$H_2CO + h\nu \rightarrow H_2 + CO$<br>$CO + OH \rightarrow CO_2 + H$<br>$H + O_2 + M \rightarrow HO_2 +M$<br>$HO_2 + O_3 \rightarrow OH + O_2 + O_2$<br>net: $CH_4 +O_3 \rightarrow H_2O+ H_2+CO_2$ |

**1g M0 Methane Loss:** Found_PAP =3.25x10$^{10}$, Shown_PAP=1.93x10$^{10}$, Total_chem = 1.24x10$^{11}$

| **Oxidation $3O_2$-a (24.4%)** | **Oxidation $CH_3OOH$-b (12.8%)** | **Oxidation $O_2$ (10.4%)** | **Oxidation $6O_2$ (6.1%)** | **Oxidation $CH_3OOH$-c (5.7%)** |
|---|---|---|---|---|
| | $CH_4 + OH \rightarrow CH_3 + H_2O$<br>$CH_3 + O_2 + M \rightarrow CH_3O_2 + M$<br>$CH_3O_2+HO_2 \rightarrow CH_3OOH+O_2$<br>$H_2O + O(^1D) \rightarrow OH + OH$<br>$2(H + O_2 + M \rightarrow HO_2 +M)$<br>$NO + HO_2 \rightarrow NO_2 + OH$<br>$2(CO + OH \rightarrow CO_2 + H)$<br>$NO_2 + h\nu \rightarrow NO + O(^3P)$<br>$O(^3P) + O_2 + M \rightarrow O_3 + M$<br>$O_3 + h\nu \rightarrow O_2 + O(^1D)$<br>net: $CH_4 + 2CO + 2O_2 \rightarrow CH_3OOH + 2CO_2$ | $CH_4 + OH \rightarrow CH_3 + H_2O$<br>$CH_3 + O_2 + M \rightarrow CH_3O_2 + M$<br>$CH_3O_2 + NO \rightarrow H_3CO + NO_2$<br>$H_3CO + O_2 \rightarrow H_2CO + HO_2$<br>$HO_2 + O_3 \rightarrow OH + O_2 + O_2$<br>$NO_2 + h\nu \rightarrow NO + O(^3P)$<br>$O(^3P) + O_2 + M \rightarrow O_3 + M$<br>net: $CH_4 + O_2 \rightarrow H_2CO+H_2O$ | | $2(CH_4 + OH \rightarrow CH_3 + H_2O)$<br>$2(CH_3+O_2+M \rightarrow CH_3O_2+M)$<br>$CH_3O_2+OH \rightarrow CH_3OOH+O_2$<br>$CH_3O_2 + NO \rightarrow H3CO + NO_2$<br>$H_3CO + O_2 \rightarrow H_2CO + HO_2$<br>$NO_2 + h\nu \rightarrow NO + O(^3P)$<br>$O(^3P) + O_2 + M \rightarrow O_3 + M$<br>$O_3 + h\nu \rightarrow O_2 + O(^1D)$<br>$H_2O + O(^1D) \rightarrow OH + OH$<br>net: $2CH_4 + 2O_2 \rightarrow H_2CO + H_2O + CH_3OOH$ |

**3g M0 Methane Loss**: Found_PAP =5.67x10$^{10}$, Shown_PAP=2.16x10$^{10}$, Total_chem = 1.24x10$^{11}$

| **Oxidation $CH_3OOH$-a (12.6%)** | **Oxidation $2O_2$-b (8.5%)** | **Oxidation $2O_2$-c (6.8%)** | **Oxidation $2O_2$-d (5.5%)** | **Oxidation $CH_3OOH$-b (4.7%)** |
|---|---|---|---|---|

**1g ADL Methane Loss**: Found_PAP =6.14x10$^{10}$, Shown_PAP=3.95x10$^{10}$, Total_chem = 1.24x10$^{11}$

| | | | | |
|---|---|---|---|---|
| **Oxidation $2O_2$-b (30.0%)** | **Oxidation $2O_2$-d (14.7%)**<br>$CH_4 + OH \rightarrow CH_3 + H_2O$<br>$CH_3 + O_2 + M \rightarrow CH_3O_2 + M$<br>$CH_3O_2 + NO \rightarrow H_3CO + NO_2$<br>$H_3CO + O_2 \rightarrow H_2CO + HO_2$<br>$NO_2 + O(^3P) \rightarrow NO + O_2$<br>$H_2CO + OH \rightarrow H_2O + HCO$<br>$HCO + O_2 \rightarrow HO_2 + CO$<br>$CO + OH \rightarrow CO_2 + H$<br>$H + O_2 + M \rightarrow HO_2 + M$<br>$3(HO_2 + O(^3P) \rightarrow OH + O_2)$<br>$2(O_2 + h\nu \rightarrow O(^3P) + O(^3P))$<br>net: $CH_4 + 2O_2 \rightarrow 2H_2O + CO_2$ | **Oxidation $2O_2$-c (7.3%)** | **Oxidation $2O_2$-e (6.2%)**<br>$CH_4 + OH \rightarrow CH_3 + H_2O$<br>$CH_3 + O_2 + M \rightarrow CH_3O_2 + M$<br>$CH_3O_2 + OH \rightarrow H_3CO + HO_2$<br>$H_3CO + O_2 \rightarrow H_2CO + HO_2$<br>$H_2CO + OH \rightarrow H_2O + HCO$<br>$HCO + O_2 \rightarrow HO_2 + CO$<br>$3(HO_2 + O(^3P) \rightarrow OH + O_2)$<br>$CO + OH \rightarrow CO_2 + H$<br>$H + O_3 \rightarrow OH + O_2$<br>$O(^3P) + O_2 + M \rightarrow O_3 + M$<br>$2(O_2 + h\nu \rightarrow O(^3P) + O(^3P))$<br>net: $CH_4 + 2O_2 \rightarrow 2H_2O + CO_2$ | **Oxidation $2O_2$-Cl (6.2%)**<br>$CH_4 + Cl \rightarrow HCl + CH_3$<br>$CH_3 + O_2 + M \rightarrow CH_3O_2 + M$<br>$CH_3O_2 + OH \rightarrow H_3CO + HO_2$<br>$H_3CO + O_2 \rightarrow H_2CO + HO_2$<br>$H_2CO + OH \rightarrow H_2O + HCO$<br>$HCO + O_2 \rightarrow HO_2 + CO$<br>$HCl + OH \rightarrow Cl + H_2O$<br>$CO + OH \rightarrow CO_2 + H$<br>$H + O_2 + M \rightarrow HO_2 + M$<br>$4(HO_2 + O(^3P) \rightarrow OH + O_2)$<br>$2(O_2 + h\nu \rightarrow O(^3P) + O(^3P))$<br>net: $CH_4 + 2O_2 \rightarrow 2H_2O + CO_2$ |

**3g ADL Methane Loss**: Found_PAP =7.06x10$^{10}$, Shown_PAP=3.31x10$^{10}$, Total_chem = 1.24x10$^{11}$

| | | | | |
|---|---|---|---|---|
| **Oxidation $2O_2$-b (21.9%)** | **Oxidation $2O_2$-c (6.3%)** | **Oxidation $2O_2$-d (6.3%)** | **Oxidation $2O_2$-e (6.2%)** | **Oxidation $2O_2$ -f (6.1%)**<br>$CH_4 + O(^1D) \rightarrow CH_3 + OH$<br>$CH_3 + O_2 + M \rightarrow CH_3O_2 + M$<br>$CO + OH \rightarrow CO_2 + H$<br>$H + O_2 + M \rightarrow HO_2 + M$<br>$CH_3O_2 + HO_2 \rightarrow CH_3OOH + O_2$<br>$CH_3OOH + OH > CH_3O_2 + H_2O$<br>$CH_3O_2 + OH \rightarrow H_3CO + HO_2$<br>$H_3CO + O_2 \rightarrow H_2CO + HO_2$<br>$H_2CO + OH \rightarrow H_2O + HCO$<br>$HCO + O_2 \rightarrow HO_2 + CO$<br>$3(HO_2 + O(^3P) \rightarrow OH + O_2)$<br>$O(^3P) + O_2 + M \rightarrow O_3 + M$<br>$O_3 + h\nu \rightarrow O_2 + O(^1D)$<br>$2(O_2 + h\nu \rightarrow O(^3P) + O(^3P))$<br>net: $CH_4 + 2O_2 \rightarrow 2H_2O + CO_2$ |

**1g M5 Methane Loss:** Found_PAP =5.65x10$^{10}$, Shown_PAP=4.04x10$^{10}$, Total_chem = 1.24x10$^{11}$

| | | | | |
|---|---|---|---|---|
| **Oxidation $2O_2$ (35.5%)**<br>$CH_4 + OH \rightarrow CH_3 + H_2O$<br>$CH_3 + O_2 + M \rightarrow CH_3O_2 + M$<br>$CH_3O_2 + NO \rightarrow H_3CO + NO_2$<br>$H_3CO + O_2 \rightarrow H_2CO + HO_2$<br>$H_2CO + OH \rightarrow H_2O + HCO$<br>$HCO + O_2 \rightarrow HO_2 + CO$<br>$CO + OH \rightarrow CO_2 + H$<br>$H + O_2 + M \rightarrow HO_2 + M$<br>$3(NO + HO_2 \rightarrow NO_2 + OH)$<br>$2(NO_2 + h\nu \rightarrow NO + O(^3P))$<br>$2(NO_2 + O(^3P) \rightarrow NO + O_2)$<br>net: $CH_4 + 2O_2 \rightarrow 2H_2O + CO_2$ | **Oxidation $3O_2$-a (12.7%)** | **Oxidation $O_2$ (11.4%)** | **Oxidation $3O_2$-b (6.7%)** | **Oxidation $2O_2$-g (5.3%)**<br>$CH_4 + OH \rightarrow CH_3 + H_2O$<br>$CH_3 + O_2 + M \rightarrow CH_3O_2 + M$<br>$CH_3O_2 + NO \rightarrow H_3CO + NO_2$<br>$H_3CO + O_2 \rightarrow H_2CO + HO_2$<br>$H_2CO + OH \rightarrow H_2O + HCO$<br>$HCO + O_2 \rightarrow HO_2 + CO$<br>$NO + HO_2 \rightarrow NO_2 + OH$<br>$2(NO_2 + h\nu \rightarrow NO + O(^3P))$<br>$2(HO_2 + O(^3P) \rightarrow OH + O_2)$<br>$CO + OH \rightarrow CO_2 + H$<br>$H + O_2 + M \rightarrow HO_2 + M$<br>net: $CH_4 + 2O_2 \rightarrow 2H_2O + CO_2$ |

**3g M5 Methane Loss**: Found_PAP =7.26x10$^{10}$, Shown_PAP=4.57x10$^{10}$, Total_chem = 1.24x10$^{11}$

| | | | | |
|---|---|---|---|---|
| **Oxidation $CH_3OOH$-d (24.5%)**<br>$2(CH_4 + O(^1D) \rightarrow CH_3 + OH)$<br>$2(CH_3 + O_2 + M \rightarrow CH_3O_2 + M)$<br>$2(CO + OH \rightarrow CO_2 + H)$<br>$2(H + O_2 + M \rightarrow HO_2 + M)$<br>$2(CH_3O_2 + HO_2 > CH_3OOH + O_2)$<br>$2(O(^3P) + O_2 + M \rightarrow O_3 + M)$<br>$2(O_2 + h\nu \rightarrow O_2 + O(^1D))$<br>$O_2 + h\nu \rightarrow O(^3P) + O(^3P)$<br>net: $2CH_4 + 2CO + 3O_2 \rightarrow$ $2CH_3OOH + 2CO_2$ | **Oxidation $CH_3OOH$-e (23.7%)**<br>$CH_4 + O(^1D) \rightarrow CH_3 + OH$<br>$CH_3 + O_2 + M \rightarrow CH_3O_2 + M$<br>$CH_3O_2 + HO_2 \rightarrow CH_3OOH + O_2$<br>$2(H + O_2 + M \rightarrow HO_2 + M)$<br>$NO + HO_2 \rightarrow NO_2 + OH$<br>$2(CO + OH \rightarrow CO_2 + H)$<br>$NO_2 + h\nu \rightarrow NO + O(^3P)$<br>$O(^3P) + O_2 + M \rightarrow O3 + M$<br>$O_3 + h\nu \rightarrow O_2 + O(^1D)$<br>net: $CH_4 + 2CO + 2O_2 \rightarrow$ $CH_3OOH + 2CO_2$ | **Oxidation $H_2O_2$-a (5.8%)**<br>$CH_4 + O(^1D) \rightarrow CH_3 + OH$<br>$CH_3 + O_2 + M \rightarrow CH_3O_2 + M$<br>$CO + OH \rightarrow CO_2 + H$<br>$H + O_2 + M \rightarrow HO_2 + M$<br>$CH_3O_2 + NO \rightarrow H_3CO + NO_2$<br>$H_3CO + O_2 \rightarrow H_2CO + HO_2$<br>$HO_2 + HO_2 \rightarrow H_2O_2 + O_2$<br>$NO_2 + h\nu \rightarrow NO + O(^3P)$<br>$O(^3P) + O_2 + M \rightarrow O_3 + M$<br>$O_3 + h\nu \rightarrow O_2 + O(^1D)$<br>$H_2CO + h\nu \rightarrow H_2 + CO$<br>net: $CH_4 + 2O_2 \rightarrow H_2O_2 + H_2 + CO_2$ | **Oxidation $H_2O_2$-b (4.5%)**<br>$CH_4 + O(^1D) \rightarrow CH_3 + OH$<br>$CH_3 + O_2 + M \rightarrow CH_3O_2 + M$<br>$CO + OH \rightarrow CO_2 + H$<br>$H + O_2 + M \rightarrow HO_2 + M$<br>$CH_3O_2 + NO \rightarrow H_3CO + NO_2$<br>$H_3CO + O_2 \rightarrow H_2CO + HO_2$<br>$HO_2 + HO_2 \rightarrow H_2O_2 + O_2$<br>$NO_2 + h\nu \rightarrow NO + O(^3P)$<br>$O(^3P) + O_2 + M \rightarrow O_3 + M$<br>$O_3 + h\nu \rightarrow O_2 + O(^1D)$<br>net: $CH_4 + CO + 2O_2 \rightarrow H_2CO$ $+ H_2O_2 + CO_2$ | **Oxidation $H_2O_2$-c (4.4%)**<br>$CH_4 + O(^1D) \rightarrow CH_3 + OH$<br>$CH_3 + O_2 + M \rightarrow CH_3O_2 + M$<br>$CO + OH \rightarrow CO_2 + H$<br>$CH_3O_2 + NO \rightarrow H_3CO + NO_2$<br>$H_3CO + O_2 \rightarrow H_2CO + HO_2$<br>$H_2CO + h\nu \rightarrow HCO + H$<br>$HCO + O_2 \rightarrow HO_2 + CO$<br>$NO_2 + h\nu \rightarrow NO + O(^3P)$<br>$O(^3P) + O_2 + M \rightarrow O_3 + M$<br>$O_3 + h\nu \rightarrow O_2 + O(^1D)$<br>$2(H + O_2 + M \rightarrow HO_2 + M)$<br>$2(HO_2 + HO_2 \rightarrow H_2O_2 + O_2)$<br>net: $CH_4 + 3O_2 \rightarrow 2H_2O_2 + CO_2$ |

**1g M7 Methane Loss**: Found_PAP =5.48x10$^{10}$, Shown_PAP=3.40x10$^{10}$, Total_chem = 1.24x10$^{11}$

| | | | | |
|---|---|---|---|---|
| **Oxidation $2O_2$-f (23.6%)** | **Oxidation $3O_2$-b (14.5%)** | **Oxidation $3O_2$-a (11.1%)** | **Oxidation $2O_2$-h (7.3%)**<br>$CH_4 + OH \rightarrow CH_3 + H_2O$<br>$CH_3 + O_2 + M \rightarrow CH_3O_2 + M$<br>$CH_3O_2 + NO \rightarrow H_3CO + NO_2$<br>$H_3CO + O_2 \rightarrow H_2CO + HO_2$<br>$CH_3O_2+HO_2 \rightarrow CH_3OOH+O_2$<br>$NO_2 + O(^3P) \rightarrow NO + O_2$<br>$CH_3OOH+OH \rightarrow CH_3O_2+H_2O$<br>$H_2CO + O(^3P) \rightarrow HCO + OH$<br>$HCO + O_2 \rightarrow HO_2 + CO$<br>$CO + OH \rightarrow CO_2 + H$<br>$H + O_2 + M \rightarrow HO_2 +M$<br>$2(NO + HO_2 \rightarrow NO_2 + OH)$<br>$\underline{2(NO_2 + h\nu \rightarrow NO + O(^3P))}$<br>net: $CH_4 +2O_2 \rightarrow 2H_2O+CO_2$ | **Oxidation $2O_2$-i (5.4%)**<br>$CH_4 + OH \rightarrow CH_3 + H_2O$<br>$CH_3 + O_2 + M \rightarrow CH_3O_2 + M$<br>$CH_3O_2 + NO \rightarrow H_3CO + NO_2$<br>$H_3CO + O_2 \rightarrow H_2CO + HO_2$<br>$NO_2 + O(^3P) \rightarrow NO + O_2$<br>$H_2CO + O(^3P) \rightarrow HCO + OH$<br>$HCO + O_2 \rightarrow HO_2 + CO$<br>$2(NO + HO_2 \rightarrow NO_2 + OH)$<br>$CO + OH \rightarrow CO_2 + H$<br>$H + O_2 + M \rightarrow HO_2 +M$<br>$H_2O_2 + OH \rightarrow HO_2 + H_2O$<br>$HO_2 + HO_2 \rightarrow H_2O_2 + O_2$<br>$\underline{2(NO_2 + h\nu \rightarrow NO + O(^3P))}$<br>net: $CH_4 +2O_2 \rightarrow 2H_2O+CO_2$ |

**Appendix 1**

Ozone column in Dobson Units (DU) for the 1g, 3g scenarios corresponding to the values plotted in Figures 4a, 4b as a function of stellar effective temperature ($T_{eff}$) (K).

| $T_{eff}$ | Column $O_3$ (DU) (1g) | Column $O_3$ (DU) (3g) |
|---|---|---|
| 5800 | 305 | 275 |
| 3800 | 239 | 158 |
| 3400 | 270 | 251 |
| 3100 | 85 | 16.6 |
| 2800 | 59 | 5.3 |
| 2500 | 32 | - |